\documentclass[twocolumn,twocolappendix,trackchanges]{aastex6}
\usepackage{multirow}
\usepackage{amsmath}
\usepackage{gensymb}
\usepackage{placeins}
\usepackage{paralist}
\usepackage{enumitem}
\usepackage{hyperref}
\usepackage{morefloats}
\usepackage{float}
\usepackage[all]{hypcap}
\usepackage{array}
\usepackage{ragged2e}
\usepackage{multirow}
\makeatletter
\newcount\my@repeat@count
\newcommand{\myrepeat}[2]{%
  \begingroup
  \my@repeat@count=\z@
  \@whilenum\my@repeat@count<#1\do{#2\advance\my@repeat@count\@ne}%
  \endgroup
}
\makeatother
\begin{document}
\title{Accurate recovery of H\,{\scriptsize i} velocity dispersion from radio interferometers}
\author{R. Ianjamasimanana\altaffilmark{1}, W.J.G. de  Blok\altaffilmark{2,3,5}, George
  H. Heald\altaffilmark{4,5}} 
  \altaffiltext{1}{Max-Planck Institut f$\rm{\ddot{u}}$r Astronomie, K\"onigstuhl 17, 69117, Heidelberg, Germany}
   \email{roger@mpia.de} 
  \altaffiltext{2}{Netherlands Institute for
  Radio Astronomy (ASTRON), Postbus 2, 7990 AA Dwingeloo, the
  Netherlands}
  \altaffiltext{3}{Astrophysics, Cosmology and Gravity Centre, Department of Astronomy, University of Cape
  Town, Private Bag X3, Rondebosch 7701, South Africa}  
    \email{blok@astron.nl}
   \altaffiltext{4}{CSIRO Astronomy and Space Science, 26 Dick Perry Avenue, Kensington WA 6151, Australia}
    \altaffiltext{5}{Kapteyn Astronomical Institute, University of Groningen, PO Box 800, 9700 AV, Groningen, The Netherlands}
   \email{George.Heald@csiro.au}
\accepted{for publication in the Astronomical Journal}   
\begin{abstract}
Gas velocity dispersion measures the amount of disordered motions of 
a rotating disk. Accurate estimates of this parameter are of the utmost 
importance because it is directly linked to disk stability and star formation. 
A global measure of the gas velocity dispersion can be inferred from the width of 
the atomic hydrogen (H\,{\sc i}) 21 cm line. We explore how several systematic 
effects involved in the production of H\,{\sc i} cubes affect the estimate of 
H\,{\sc i} velocity dispersion. We do so by comparing the H\,{\sc i} velocity 
dispersion derived from different types of data cubes provided by The H\,{\sc i} 
Nearby Galaxy Survey (THINGS). We find that residual-scaled cubes best recover the 
H\,{\sc i} velocity dispersion, independent of the weighting scheme used and for a 
large range of signal-to-noise ratio. For H\,{\sc i} observations where the dirty beam is 
substantially different from a Gaussian, the velocity dispersion values are overestimated 
unless the cubes are cleaned close to (e.g., $\sim$1.5 times) the noise level.   
       
\end{abstract}
\keywords{}
\section{Introduction}
The random motion of gas in galaxies can be traced from the width of 
the ubiquitous 21-cm line of atomic hydrogen (H\,{\sc i}). The amount 
of the disordered motion, the gas velocity dispersion, 
provides valuable information on the physical properties of the gas 
such as its ability to collapse and form stars, the energy balance and 
the phase structure of the gas \citep{schaye04, tamburroetal09, obrienetal10, 
stilpetal13, ianjamasimananaetal12}. This parameter is usually 
determined by fitting H\,{\sc i} profiles with an assumed model, e.g., 
a single or a double Gaussian function, 
a Gauss-Hermite polynomial, or a Lorentzian 
\citep{braun97, youngetal03, ianjamasimananaetal12}. Instead of assuming a model, 
the intensity-weighted standard deviation of the line-of-sight 
velocities has also been used as a measure of 
H\,{\sc i} velocity dispersion \citep[e.g.,][]{tamburroetal09}. 

An accurate determination of the H\,{\sc i} velocity dispersion is limited 
by the ability of the imaging process to recover the intrinsic shapes 
of the H\,{\sc i} line profiles. In radio interferometry, the source intensity distribution is not completely sampled 
by the telescope array. Therefore, to recover the missing values in the $(u, v)$ data and reconstruct the 
true image of the source, the telescope Point Spread Function (PSF or dirty beam) needs to be 
deconvolved from the observed (dirty) image. As will be explained in detail later, 
the recovered image from a deconvolution algorithm is obtained by 
adding a deconvolved map of the model of the source brightness distribution (derived down to a certain flux) to 
a map containing the remaining emission, usually assumed to be noise (the residual map). 

The width of 
the H\,{\sc i} velocity profiles means that typically they will be spread over several velocity channels. 
The faint outer wings of these profiles can thus appear in adjacent channels but with intensities 
below the threshold for cleaning. The residual image will then still contain faint 
residual emission corresponding to these wings. In this paper, we investigate 
what the effect of this uncleaned, residual emission is on measurements of the 
velocity dispersion and find that it leads to an overestimate of this parameter. 
Here we focus on data obtained as part of the THINGS survey \citep{walter08}, 
but our conclusions will apply to all H\,{\sc i} observations where the dirty beam is 
substantially different from a Gaussian, such as is the case with multi-configuration 
Very Large Array (VLA) observations. This includes recent surveys such as, 
Local Irregulars That Trace Luminosity Extremes: 
The H\,{\sc i} Nearby Galaxy Survey \citep[Little THINGS;][]{hunteretal12}, 
VLA survey of Advanced Camera for Surveys Nearby Galaxy Survey Treasury 
galaxies \citep[VLA-ANGST;][]{ott12}, Survey of H\,{\sc i} in Extremely 
Low-mass Dwarfs \citep[SHIELD;][]{cannonetal11}. This study is also 
relevant for other arrays that produce a non-Gaussian dirty beam, 
e.g., the Atacama Large Millimeter/submillimeter Array (ALMA), meaning the 
results could also be relevant for molecular line velocity dispersions. 

In Section~\ref{sec:background}, we give background on the weighting of visibilities and cleaning processes. 
In Section~\ref{sec:goal}, we describe the analysis goal and method.   
In Section~\ref{sec:1}, we compare H\,{\sc i} velocity dispersion from different types of data cubes. 
In Section~\ref{sec:2}, we investigate 
the type of data cubes that should be used in H\,{\sc i} velocity dispersion analysis. 
In  Section~\ref{sec:discussion}, we discuss our results and conclusions. 
\section{Background}\label{sec:background}
Interferometric observations of radio astronomical sources provide noisy samples of the complex visibility function of the sources 
at discrete locations in the $(u,v)$ plane \citep{cornwell08}, where $u$ and $v$ are the baseline vectors projected onto a plane perpendicular to the source direction. 
Deconvolution algorithms attempt to recover the image information (missing values) 
in the un-sampled $(u,v)$ plane. The inverse Fourier transform of the sampled visibility function, $S(u, v)~V(u, v)$, is called \textit{the dirty image}, $I_{D}$: 
\begin{equation} \label{eq:num1}
I_{D}=\mathcal{F}^{-1}(S(u, v)~V(u, v)),
\end{equation}
where $S(u, v)$ is a sampling function and $V(u, v)$ is the visibility function.  
Using the Fourier convolution theorem, which states that the Fourier transform of the product of two functions is the 
convolution of the Fourier transform of the two functions, Equation~\ref{eq:num1} becomes: 
\begin{equation} \label{eq:DirtyF}
I_{D}=\mathcal{F}^{-1}(S(u, v)) \otimes \mathcal{F}^{-1}(V(u, v)). 
\end{equation}    
In the radio astronomy jargon, $\mathcal{F}^{-1}(S(u, v))$ is called \textit{the dirty beam}. The true image is the 
inverse Fourier transform of the visibility function, $\mathcal{F}^{-1}(V(u, v))$. A deconvolution algorithm 
involves solving for the true image 
given the dirty beam and the dirty image. To get the dirty image from the sampled visibility functions, 
a Fast Fourier Transform (FFT) is usually preferable over the \textit{simple} Fourier Transform (FT) due to its 
computational speed. However, FFT requires that the sample visibilities are on a regularly spaced grid (cell) 
\citep[e.g.,][]{yatawatta14}. 
In the image reconstruction and gridding process, a weighting function is usually applied to the sampled visibility function to, e.g., 
maximise point source sensitivity or to minimise sidelobe levels. If $W(u, v)$ is a weighting function, then 
Equation~\ref{eq:num1} becomes:
\begin{equation}\label{eq:weighting}
I_{D}=\mathcal{F}^{-1}(W(u, v)~S(u, v)~V(u, v)).
\end{equation}
Many forms of weighting functions have been developed for different science goals. 
This paper discusses the effects of using different weighting schemes in the derivation of H\,{\sc i} 
velocity dispersion. Natural, Uniform and Robust (or Briggs) weightings are widely 
known weighting schemes in radio interferometry. 
The Natural weighting gives a weight of 1/$\sigma_{i}^{2}$ on visibility $i$, 
where $\sigma_{i}^{2}$ is the noise variance of that 
visibility. In most cases, the noise is roughly similar for all visibilities 
and thus each $(u,v)$ points gets approximately the same weight. 
Due to the gridding, $(u,v)$ cells at shorter baselines get more 
weight as the inner parts of the $(u, v)$ plane have a higher density 
of $(u, v)$ points (i.e., measured visibilities). Thus, Natural 
weighting  offers the best surface brightness sensitivity and minimum noise, 
but at the cost of a larger beam size.  
For Uniform weighting, a given cell in the $(u,v)$ plane gets a weight 
of 1/$\rho_{(u, v)}$, where $\rho_{(u, v)}$ is the density of the $(u,v)$ 
points in the cell. Given that the density of the $(u,v)$ 
points is usually lower at larger baselines, the latter are weighted more, 
resulting in better resolution but lower sensitivity (short 
baselines are de-emphasized), and  therefore an increased rms noise. 
The Uniform weighting gives better (i.e., lower) side lobes than the 
Natural weighting. Robust weighting combines the advantage offered by the Natural weighting and the Uniform weighting. 
It gives a resolution approaching that of a Uniform weighting, with only a modest cost in sensitivity \citep{briggs95}. 
After gridding the sample visibilities and performing an FFT on Equation \ref{eq:weighting}, the obtained dirty image 
needs to be deconvolved to get an estimate of the true image. CLEAN, first described by \citet{hogbom74}, 
is among the widely used algorithms for this purpose. There have been many variants of CLEAN since its inception, 
most of them aiming to improve computational efficiency and performance with respect 
to extended structure. The original CLEAN algorithm, also called classical CLEAN or 
delta function CLEAN, assumes that the true image brightness distribution can be 
well represented by a collection of point sources. The algorithm iteratively searches 
for the locations and strengths of the 
point sources and iteratively subtracts them using the shape of the dirty beam to 
generate a \textit{residual map} and a \textit{clean component list}. 
It is common practice to convolve the clean component list with an idealized 
``CLEAN beam'' \citep[e.g., usually a Gaussian with a FWHM matching that of the central 
component of the dirty beam][]{richetal08}. 
The convolved clean component list is then added to the residual map to obtain the 
final restored image, which is a plausible representation of the true image.  Thus, 
the CLEAN-ed map  is a combination of two maps, imaged with two different beams. 
This paper explores the effects of this on measurements of the H\,{\sc i} velocity dispersion.
\section{Goal and Method}\label{sec:goal}
\subsection{Residual-scaling}\label{sec:res_scale}
\begin{figure*} 
\centering
\begin{tabular}{l l}
\hspace*{0cm}
\includegraphics[scale=.34]{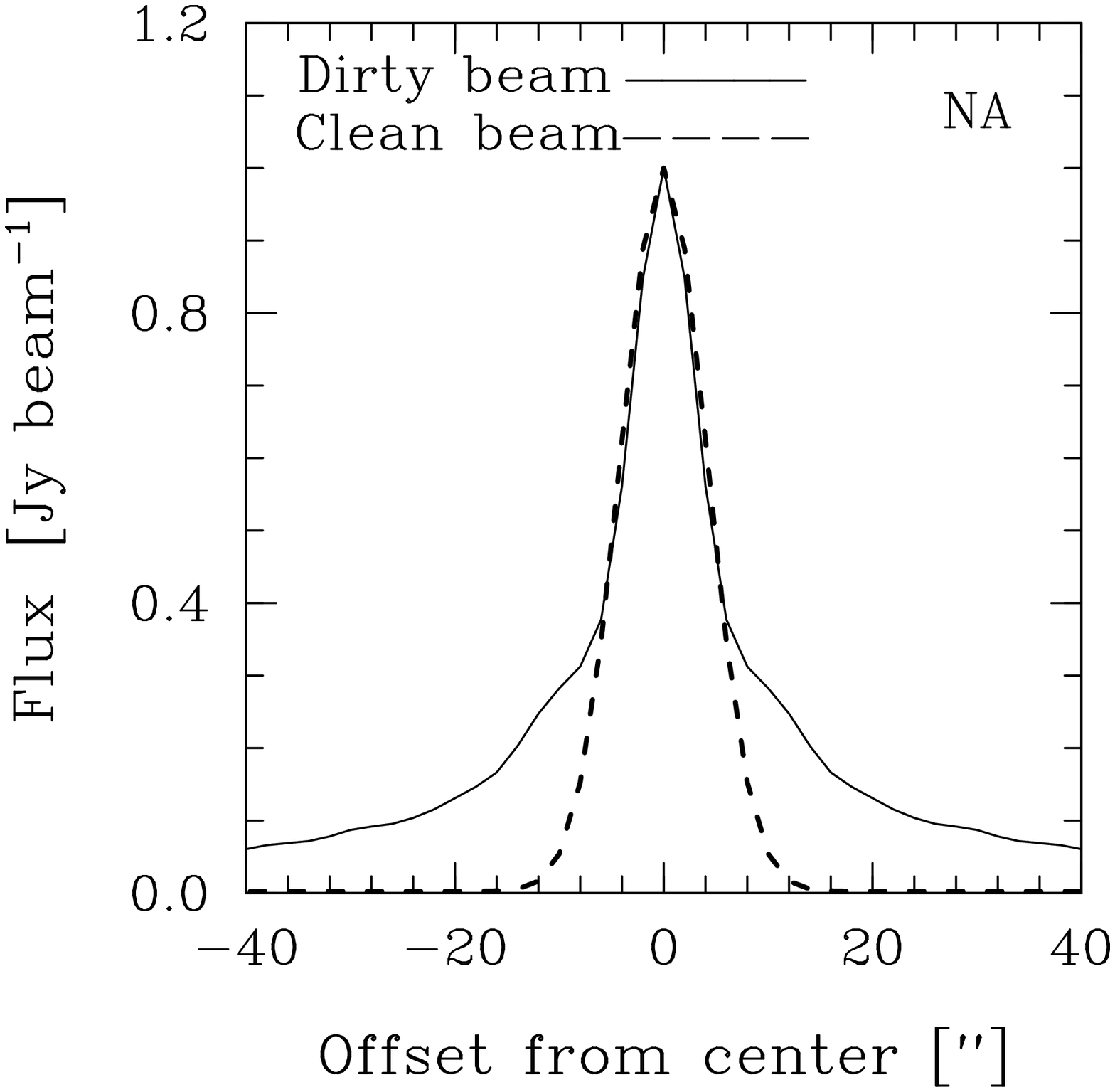} &
\hspace{1cm}
\includegraphics[scale=.34]{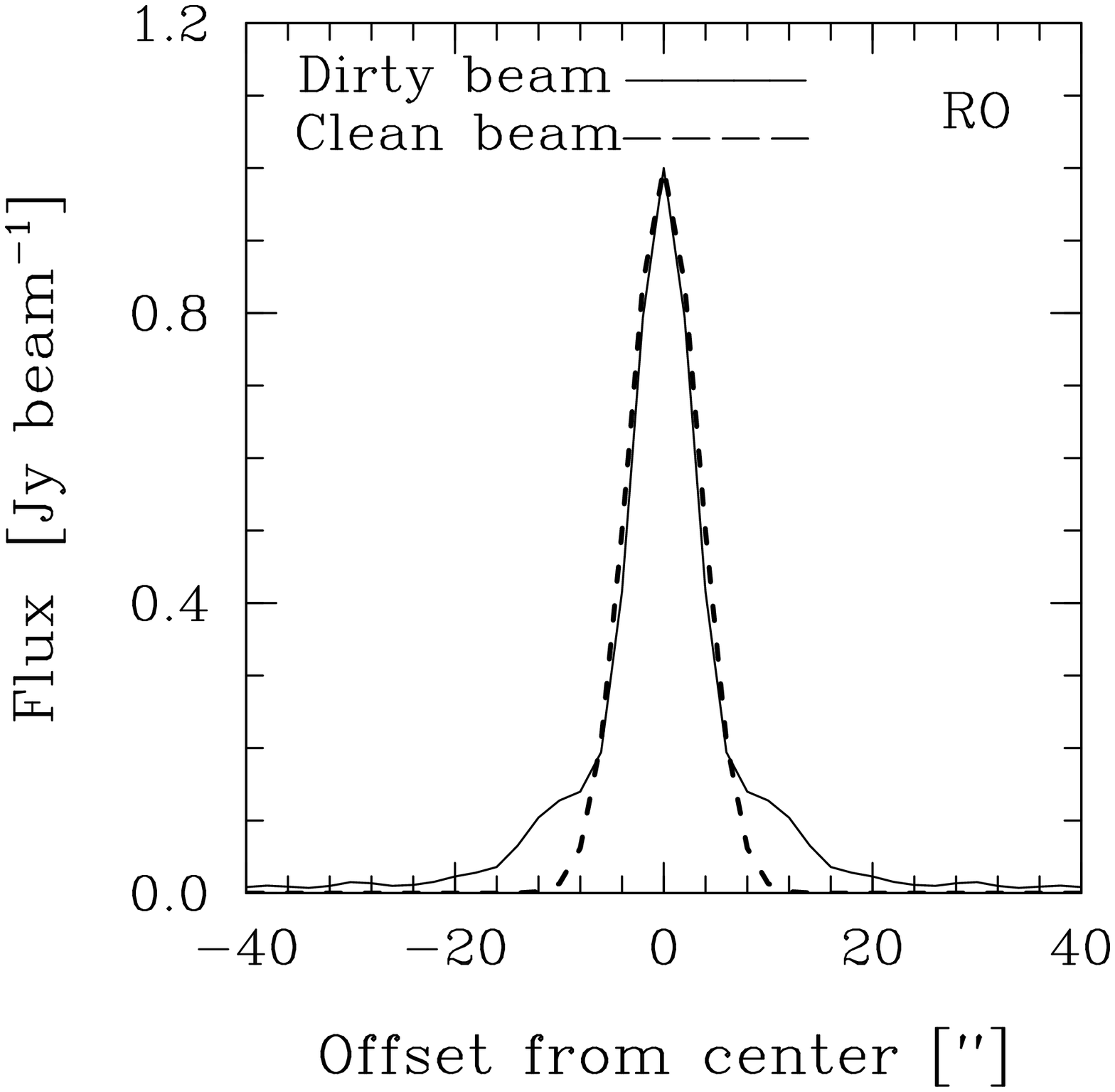}
\end{tabular}
\caption{Dirty and clean beams of the THINGS Natural (left panel) and the Robust (right panel) 
weighted data cubes of NGC 3184.}
   \label{fig:crosscut} 
\end{figure*}

Naively one would expect that the flux of a channel map is obtained by simply adding 
the flux of the cleaned map to that of the residual map:  
\begin{equation}\label{units}
F = C + R, 
\end{equation}
where $C$ is the cleaned flux and $R$ is the residual flux. However, $C$ and $R$ have different 
units ($C$ is in Jy/clean beam and $R$ in Jy/dirty beam) and depending on the configuration of the 
array, the shape of the dirty beam can be strongly non-Gaussian and more extended than the clean beam. 
This mismatch in shape between the dirty beam and the clean beam leads to an overestimate of the 
flux density if the clean beam parameters are used to determine the flux of sources containing 
both cleaned and residual emission. We illustrate in Figure~\ref{fig:crosscut}, how the dirty 
beam and the clean beam differ from each other in typical multi-configuration observations 
with the VLA. To account for the difference in area between the clean beam and the dirty beam, 
the flux of the residual must be scaled by a scaling factor $\epsilon$: 
\begin{equation}
F(true) = C + \epsilon~ R,     
\end{equation}
where $F(true)$ is the corrected flux and $\epsilon$ is a correction factor that 
accounts for the ratio between the clean beam area and the dirty beam area  \citep{jorsater95}.
Theoretically, due to the absence of zero spacing in interferometric data, 
the dirty beam integral and the total flux density of the dirty image are zero as there are 
equal amounts of positive and negative flux in the dirty image. In practice, the positive flux 
is mostly concentrated in the main lobe of the dirty beam and, for the purposes of residual-scaling, 
one has to choose a region (box) centered on the main lobe over which to integrate the 
dirty beam and ensure most of the flux in the main lobe is captured. 
The residual scaling method is included in the AIPS\footnote{The Astronomical 
Image Processing System (AIPS) has been developed by
the National Radio Astronomy Observatory (NRAO).} task IMAGR and more technical aspects are presented there. 
As described in \citet{walter08}, 
the calculation of the flux in the main lobe of the THINGS dirty beam 
was done inside a box with a half-width of 50 pixels 
(75$\arcsec$) in R.A. and in Dec. This box was chosen as it encompasses the larger part of the 
main lobe of the THINGS dirty beam while still well within its first negative sidelobe. 
This choice is appropriate for all of the THINGS galaxies as they 
have all been observed in the same manner and with the same set-up, 
meaning their dirty beams are all very similar. A quantitative analysis of the uncertainties 
associated with this choice of box size and with the use of a 
different dirty beam area has not been performed, and is beyond the scope of this paper. 
However, initial testing prior to the reduction of the THINGS data as presented in 
\citet{walter08} indicates that residual-scaling results do not 
critically depend on the box size (F. Walter, priv. com.) 
Note that the residual scaling affects the noise and thus it should only be used for flux measurements in areas 
with genuine emission \citep{walter08, richetal08, ott12}. For this reason, 
the THINGS residual-scaled cubes are blanked in areas judged to be devoid of emission. 
The criteria used by THINGS is that genuine emission must be present in at least 
three consecutive channels at or above a level of 2$\sigma$ in standard cubes 
convolved to 30\arcsec~resolution. These cubes are then used as masks to blank areas containing only noise.  
While residual-scaling has been widely used to get more accurate flux values 
\citep[e.g.,][]{walter08}, the effects its presence or absence has on H\,{\sc i} velocity dispersion have not 
yet been well explored. \citet{stilpetal13} first noticed the difference between H\,{\sc i} velocity dispersion derived 
from flux-rescaled cubes and standard cubes obtained using multi-configuration VLA observations. 
This paper makes a thorough comparison of H\,{\sc i} velocity dispersions derived 
from residual-scaled cubes and non-residual-scaled cubes from THINGS. 
\subsection{Cleaning depth}
The widely used classical CLEAN algorithm iteratively deconvolves images 
until the peak residual flux reaches an adopted threshold value. Most major 
H\,{\sc i} surveys of nearby galaxies (e.g., THINGS, Little THINGS, VLA-ANGST) 
have adopted a cleaning level of a 2.5 times the rms noise to prevent noise 
spikes from being cleaned and to avoid divergence of the CLEAN process. The effects 
of residual emissions due to the extended wings of the dirty beam can be minimized by 
deeply cleaning the data but this can be computationally expensive and could 
introduce additional artefacts. Again the choice of the cleaning depth is driven 
by obtaining a reasonable flux estimate but its effects on the measurement of 
H\,{\sc i} velocity dispersion have not been quantified previously. This analysis 
explores how cleaning depth affects H\,{\sc i} velocity dispersion and investigates a number of ways that can lead to a 
more accurate determination of the velocity dispersion.
\newpage
\subsection{Weighting of visibilities}
The choice of weighting scheme in deconvolution algorithms can also 
affect the measurement of H\,{\sc i} velocity dispersion. 
This is because the dirty beam, which depends on the configuration 
of the array and the weighting function used, 
influence the shape of the clean restoring beam and 
consequently that of the individual velocity profiles in the final restored image.
We analyze the effects of weighting of uv-data during image 
construction on H\,{\sc i} velocity dispersion by comparing velocity dispersion from   
Natural and Robust weighted data cubes. Finally, using model data cubes, 
we investigate which type of data cubes should ideally be used for velocity dispersion analysis, 
or more broadly, analysis of profile widths and shapes. In this paper 
we focus on using the THINGS data, but note that the conclusions are generally applicable to any observation 
where the dirty beam differs substantially from a Gaussian. This also implies 
that for arrays where the dirty beam is close to Gaussian 
(e.g., the Westerbork Synthesis Radio Telescope, WSRT) the need 
for residual-scaling will be much reduced or absent.  
\subsection{Sample and method}
This analysis uses 22 galaxies from THINGS. These were selected 
to be minimally affected by interactions, projection effects or major bulk motions. 
For further discussion, we refer the reader to \citet{ianjamasimananaetal12}. 
Their names and observational properties are shown in Table \ref{tab:sample} 
\citep[as also given in][]{walter08}. The data provided by THINGS include both 
Natural and Robust weighted data cubes (with a robust parameter value of 0.5). 
For each weighting, the 
non-residual scaled versions are publicly available on the THINGS website 
(\url{http://www.mpia.de/THINGS/Data.html}). The residual-scaled ones are 
available upon request. Our aim is to characterize the values of the velocity 
dispersions with high precision. So, as in  \citet{ianjamasimananaetal12}, 
individual line profiles with peak amplitude higher than 4 times 
the rms noise level are aligned to the same reference velocity and 
summed to derive what we call \textit{super profiles}. 
We use a velocity field to define the amount of shift per pixel 
to end up at the same reference velocity.   
We decompose the high S/N super profiles into narrow and 
broad Gaussian components and compare the fitted parameters 
(velocity dispersion, ratio of linestrength or flux ratio  
between the narrow and broad components) from the 
different THINGS cubes to assess the effects of deconvolution and 
residual-scaling on H\,{\sc i} profile shapes. The narrow component represents 
high brightness regions whereas the broad component contains faint 
low-level emission where the residual emission contributions can become important.
We also describe the results obtained from single Gaussian fits, 
as generally used in the literature to obtain a more global measure of the 
dispersion. We remove the effect of missing short spacings, 
which is inherent to interferometric data and manifests as negative wings in the super profiles, 
by (simultaneously) including a polynomial background or baseline in the super 
profile fits. Most of our sample galaxies require a zero or constant background whereas 
in a few cases higher order polynomial fits are needed. 
  
\section{Comparing H\,{\scriptsize i} velocity dispersion from different types of data cubes}\label{sec:1}
\subsection{Natural vs Robust  cubes}
\begin{figure*}
    \centering
    \begin{tabular}{c c c}
    \includegraphics[scale=.27]{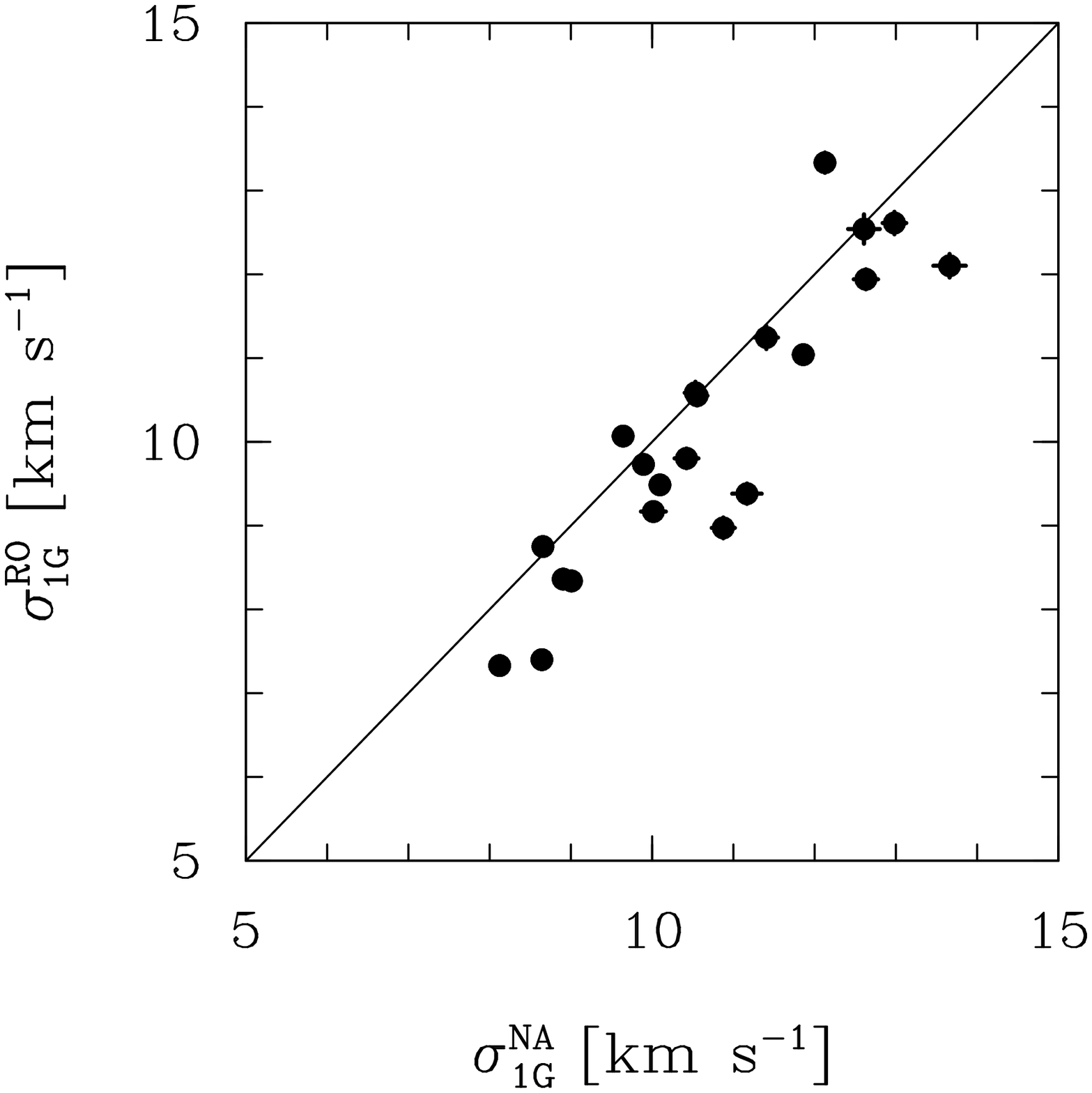}&

     \includegraphics[scale=.27]{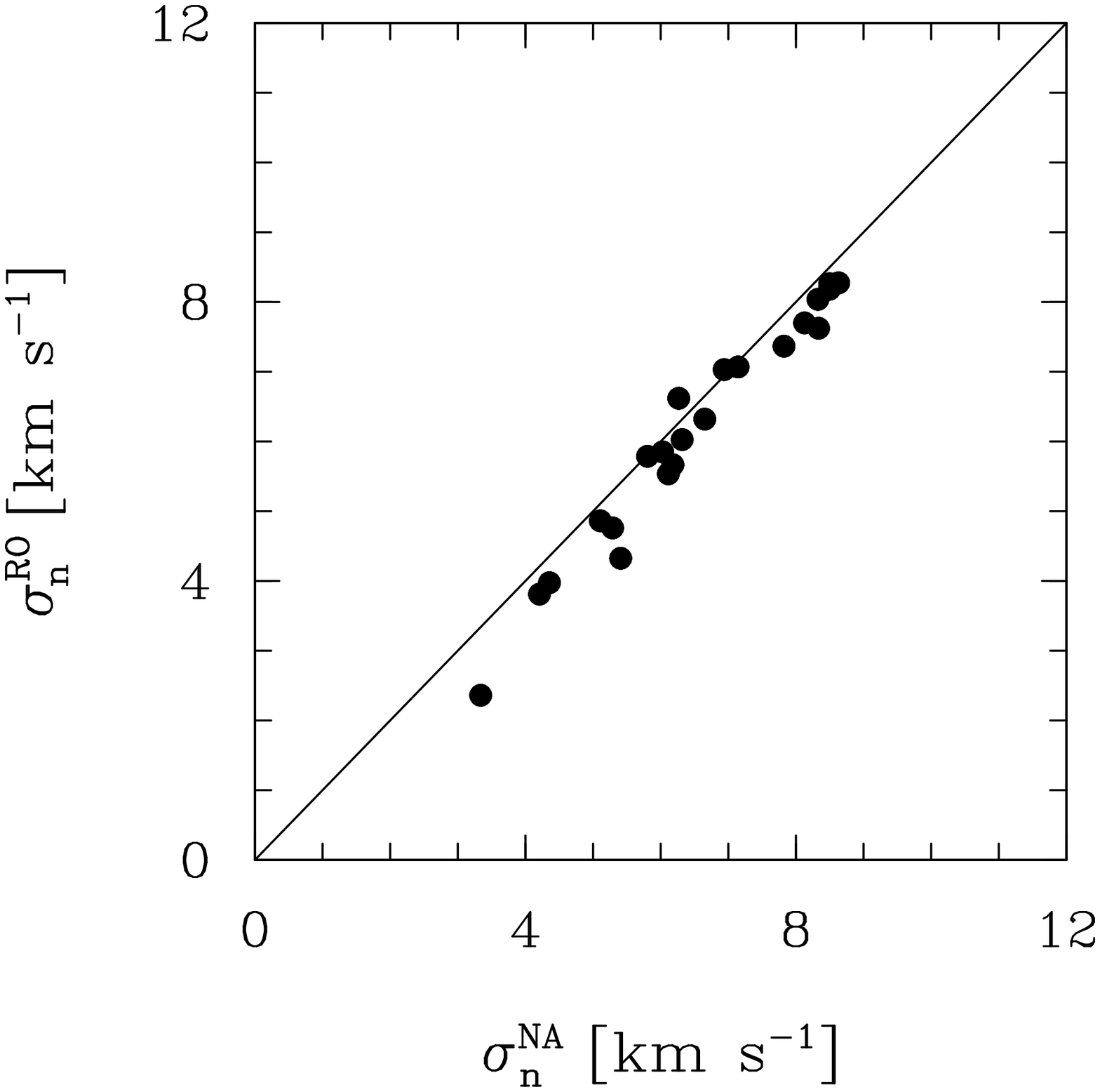}&

     \includegraphics[scale=.27]{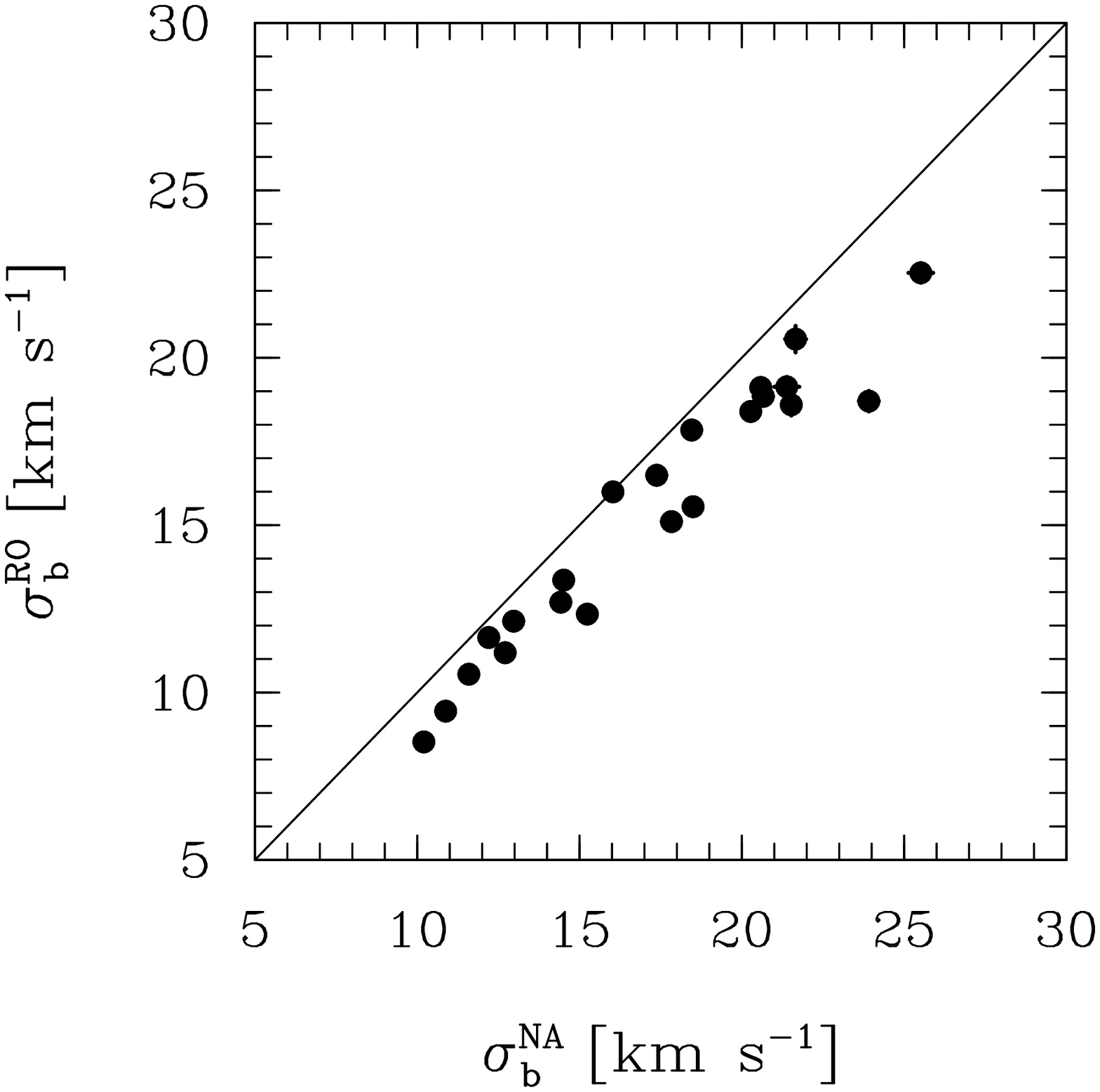}\\[.1in]
     \includegraphics[scale=.27]{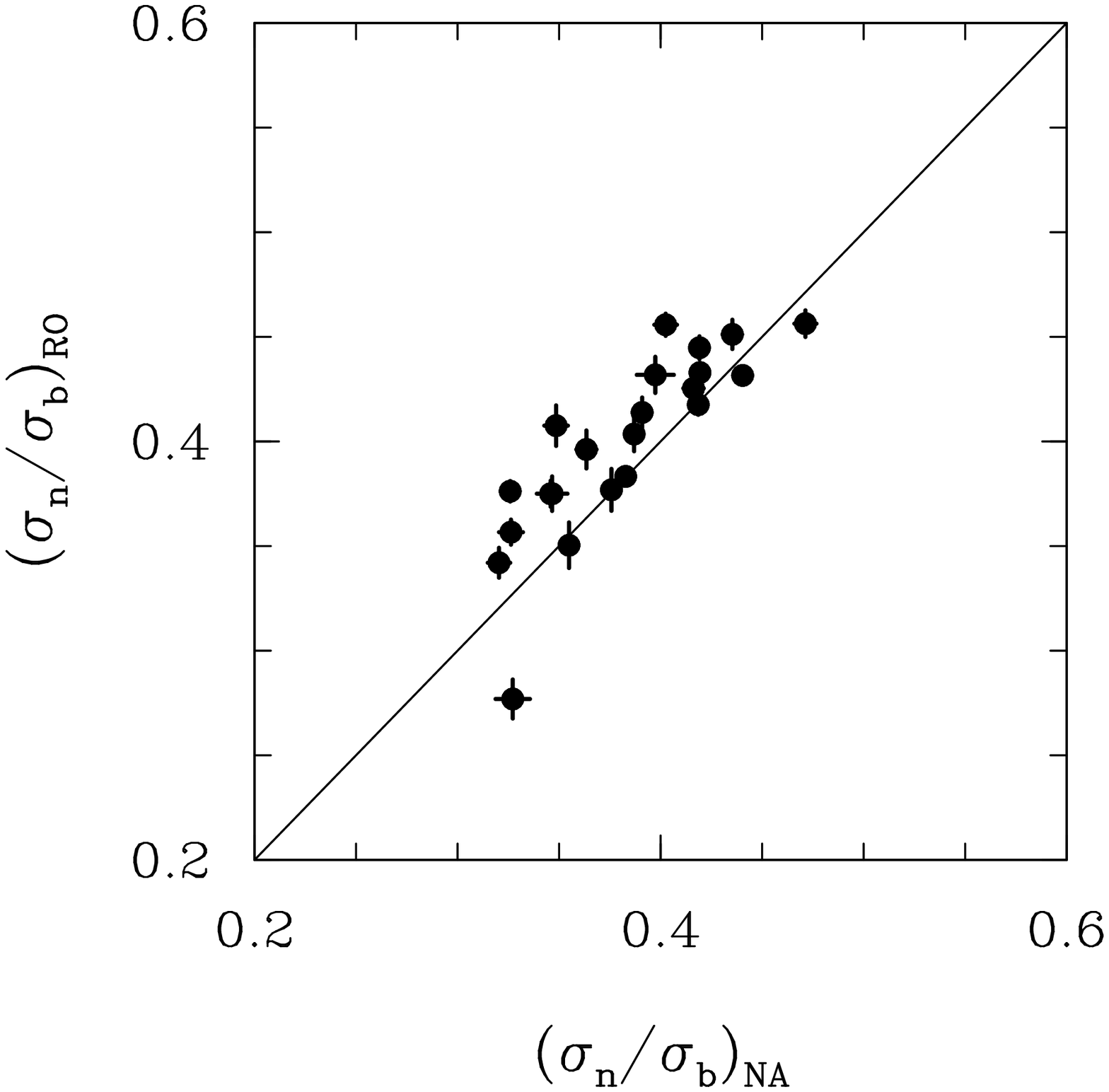}&
 
     \includegraphics[scale=.27]{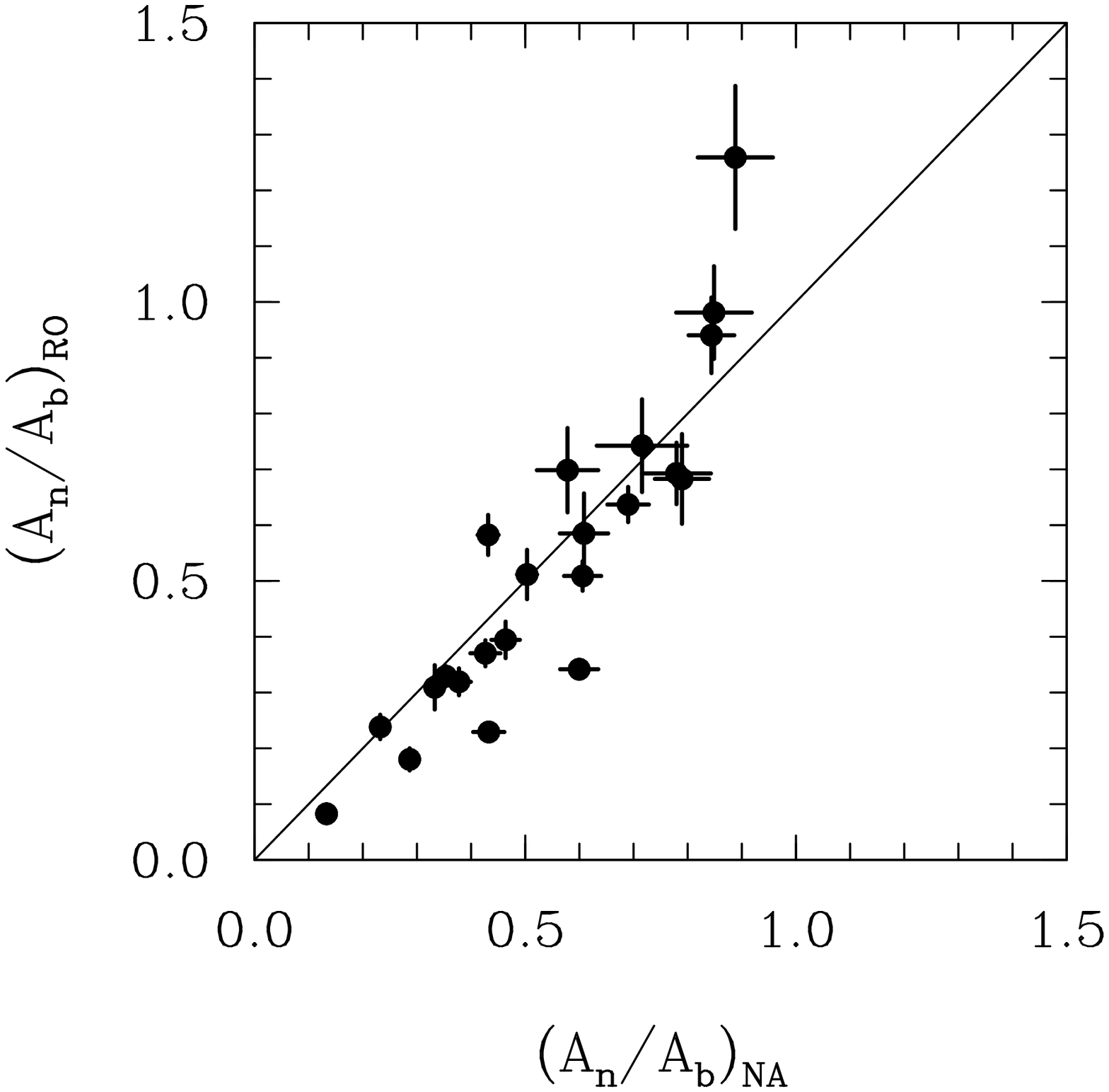}
    \end{tabular}
    \caption{Comparison of super profile parameters from natural non-residual scaled (NA) and robust non-residual scaled 
    (RO) data cubes; $\sigma_{1G}$: velocity dispersion from a single Gaussian fit; $\sigma_{n}$: 
    narrow component velocity dispersion; $\sigma_{b}$: broad component velocity dispersion; $A_{n}/A_{b}$: ratio of 
    linestrength (or flux ratio) between the narrow and broad components.} 
    \label{fig:comp_nanonres_rononres}
\end{figure*}

\begin{figure*}
\centering
    \begin{tabular}{c c c}
     \includegraphics[scale=.27]{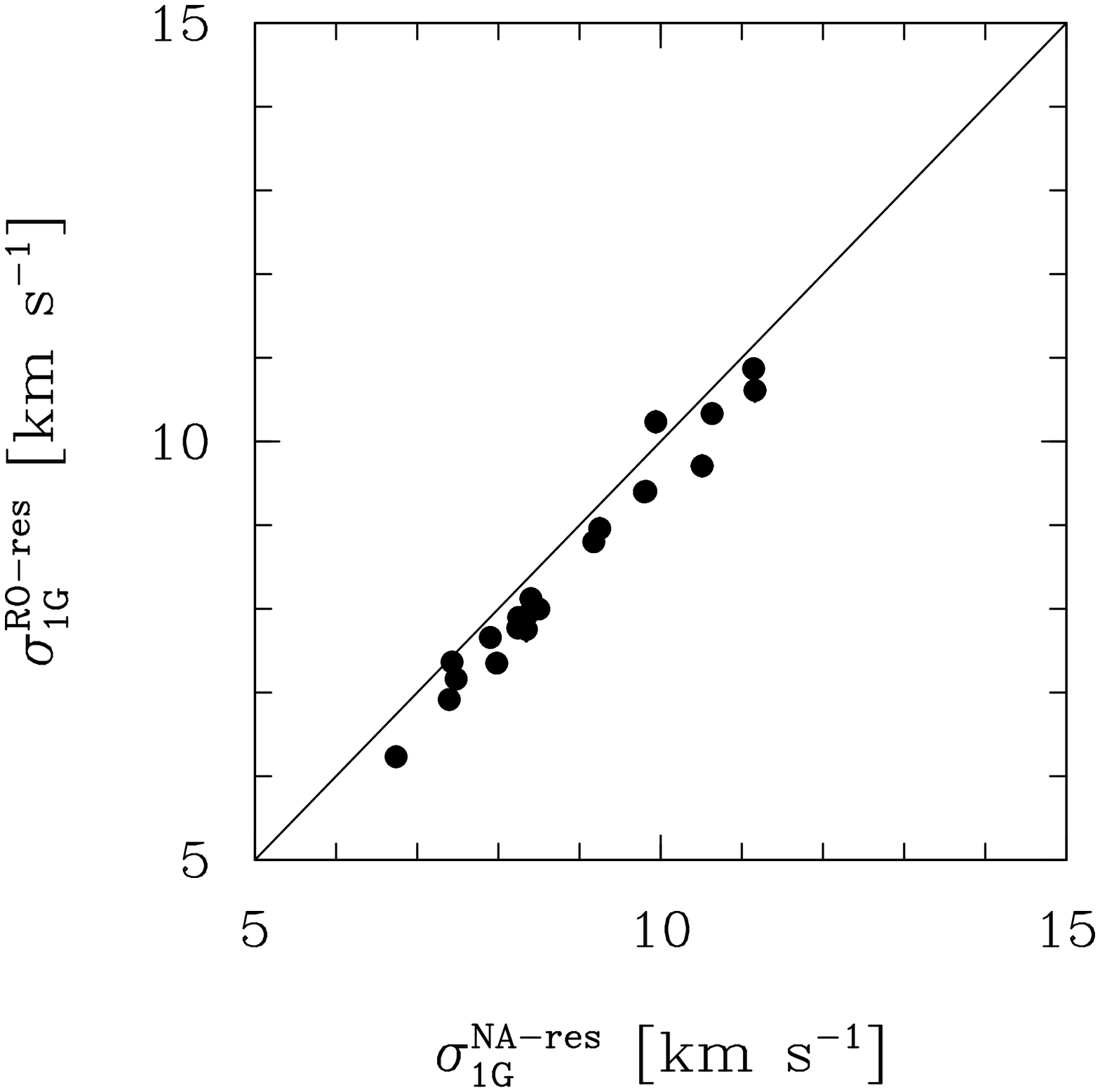}&
    
     \includegraphics[scale=.27]{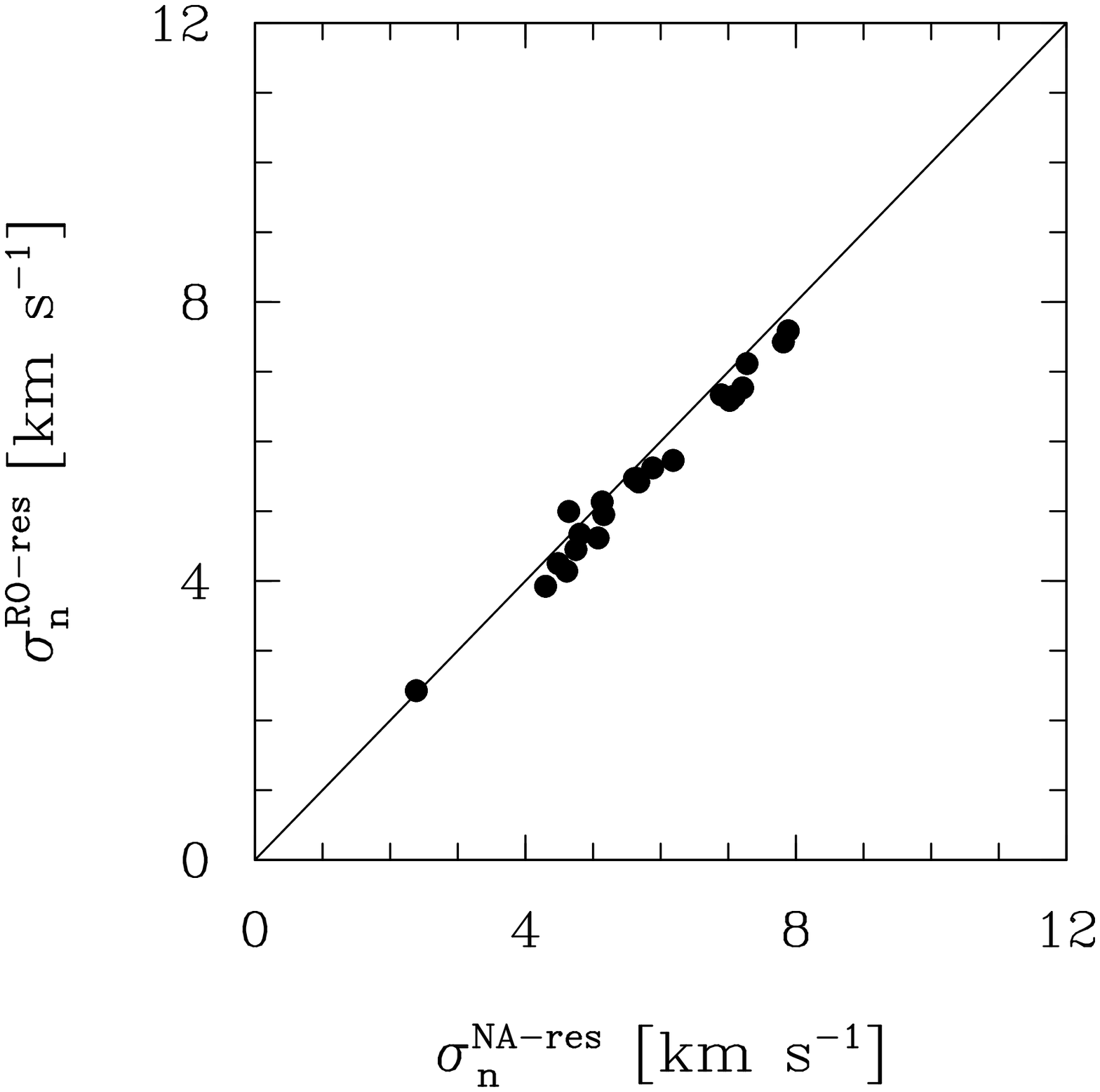}& 
  
     \includegraphics[scale=.27]{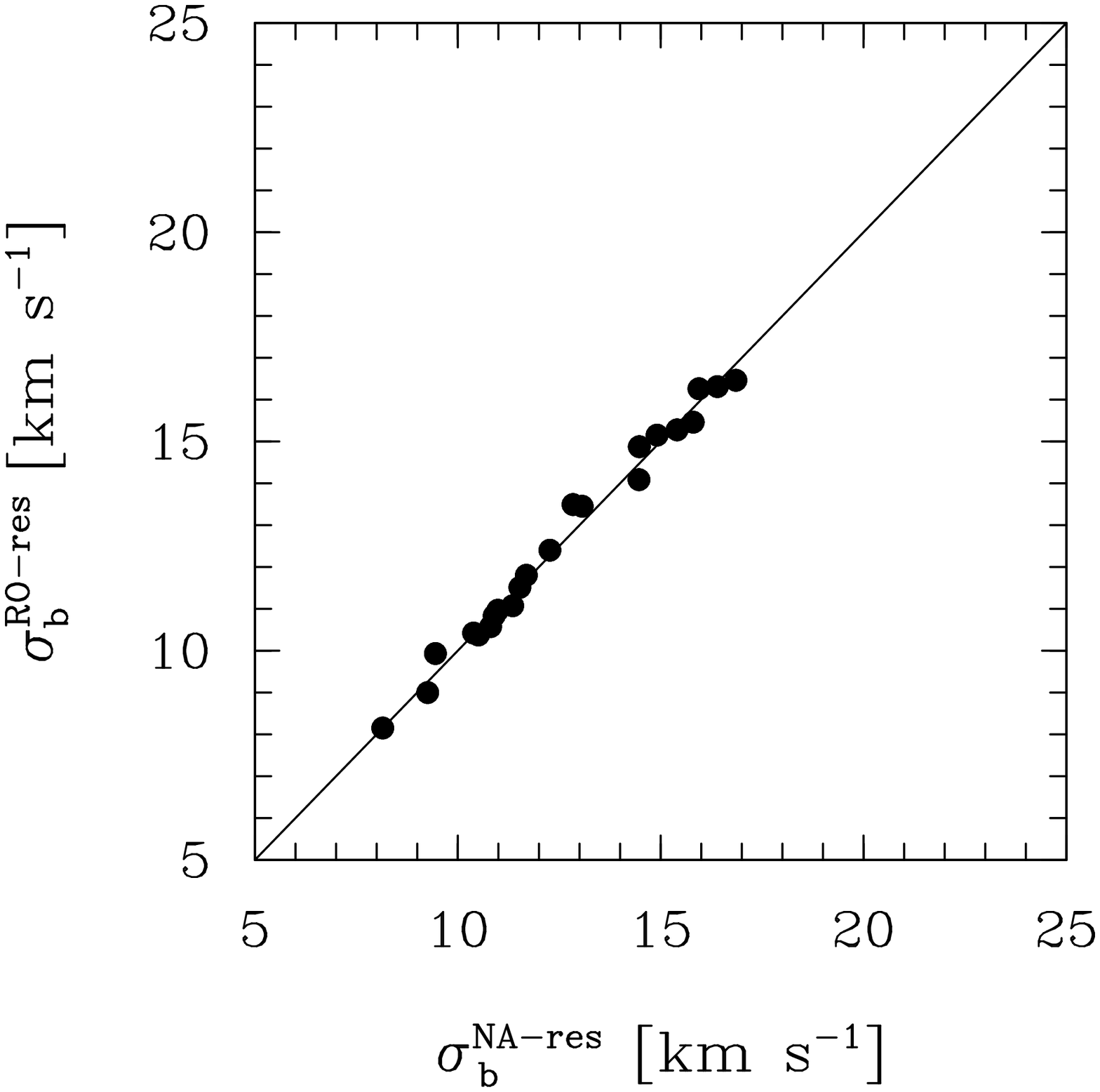}\\[.1in]
     \includegraphics[scale=.27]{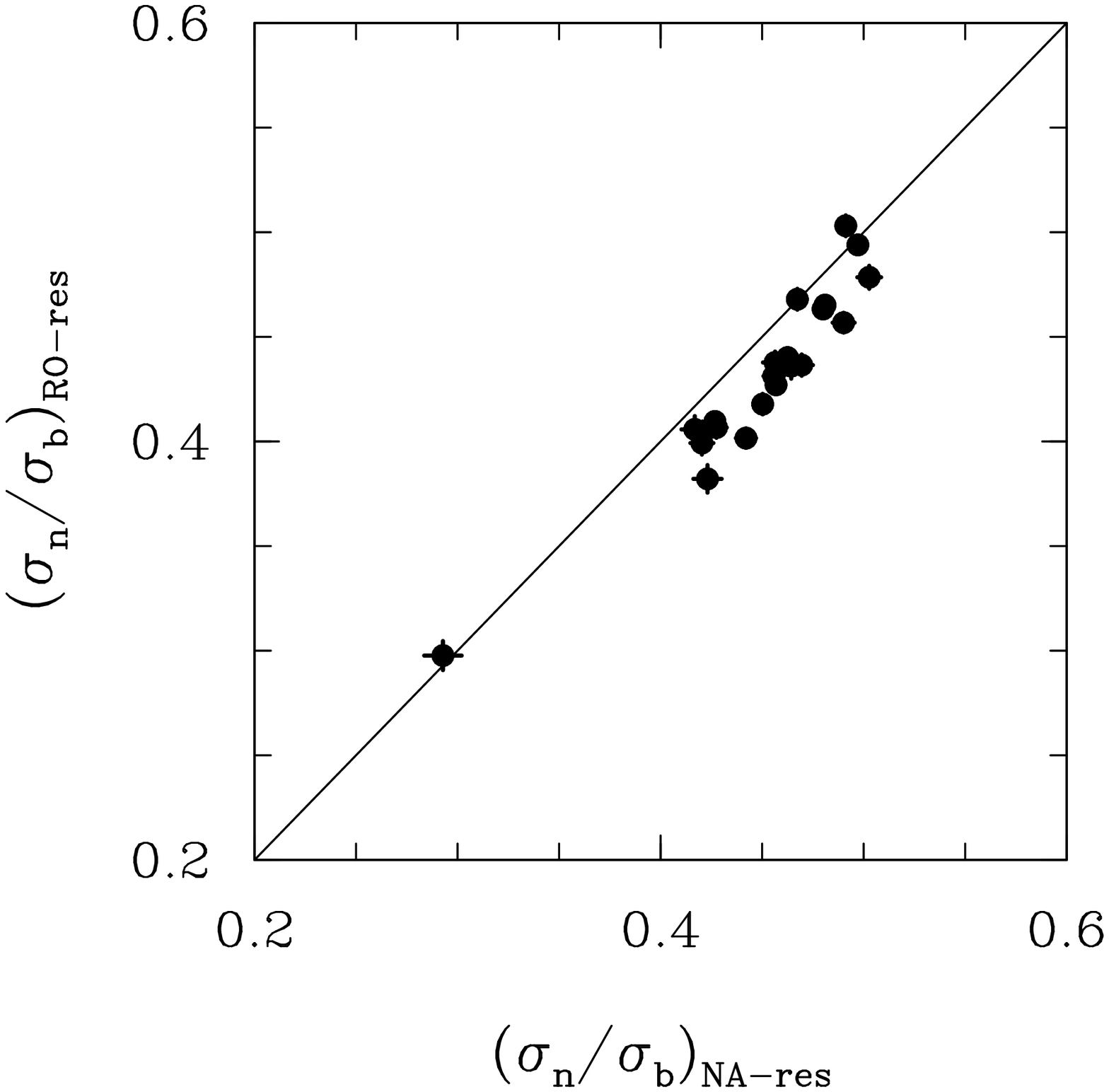}&

     \includegraphics[scale=.27]{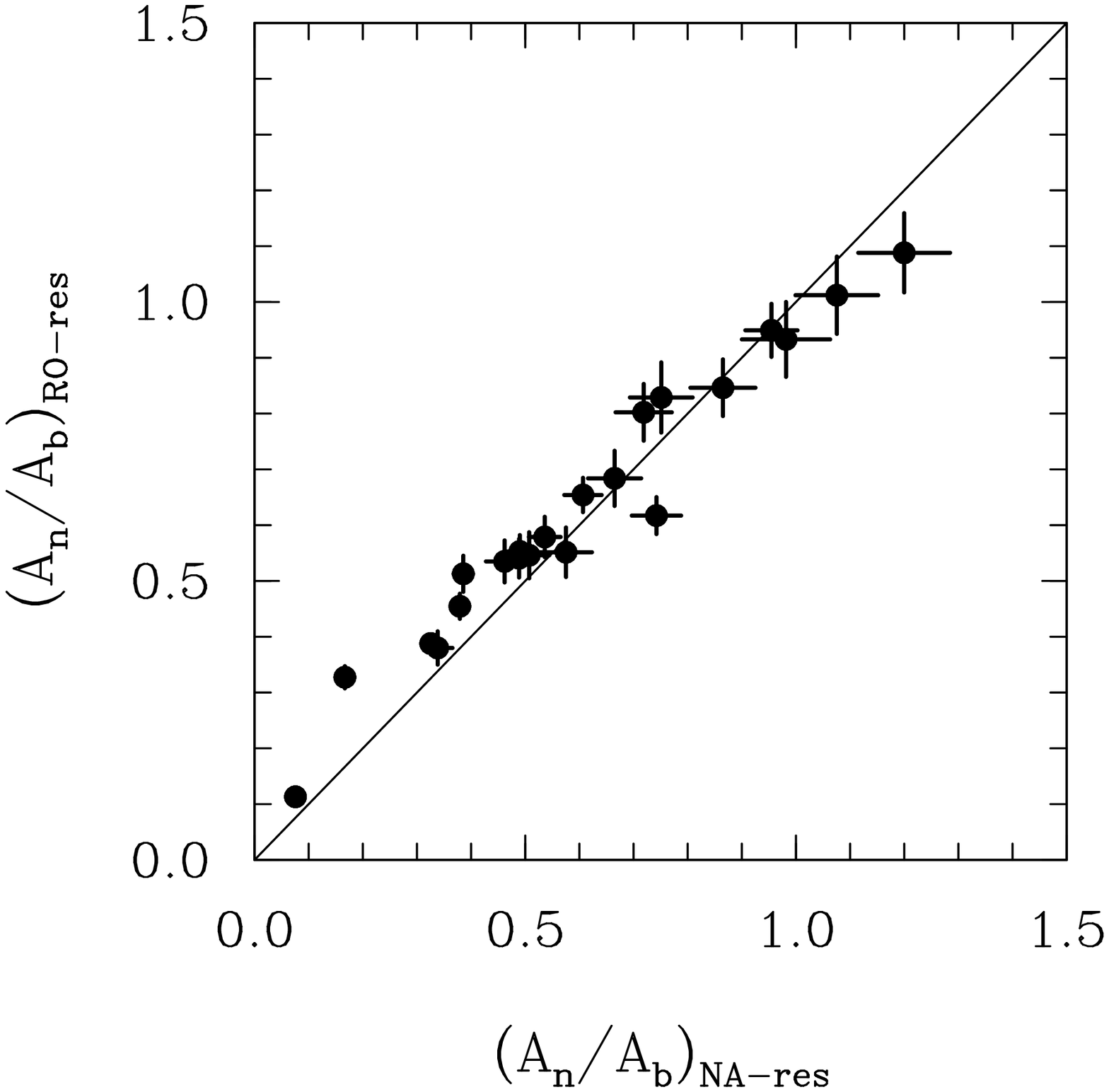}
    \end{tabular}
    \caption{Comparison of super profile parameters from natural residual-scaled (NA-res) and robust residual-scaled (RO-res) data cubes.} 
    \label{fig:comp_nares_rores}
\end{figure*}

\begin{justify}
Here we compare the super profile parameters of the Natural and the Robust THINGS data cubes. 
Figure~\ref{fig:comp_nanonres_rononres} compares the parameters from the non-residual scaled version of 
the cubes. In this case, the Natural cubes tend to give higher velocity dispersion 
values than the Robust cubes ($\sim$~7\% higher for the single Gaussian component, 
$\sigma_{1G}$, and for the narrow component, 
$\sigma_{n}$, and $\sim$~13\% higher for the broad component, $\sigma_{b}$).  
The ratio of linestrength or ratio between the narrow and broad component fluxes, 
$A_{n}/A_{b}$, however, only scatters around the line of equality. Figure~\ref{fig:comp_nares_rores} 
presents a comparison of the super profile parameters for the 
residual-scaled version of the cubes. In this case, the parameters from the 
Natural and the Robust cubes are very similar. 
We conclude that the Robust and the Natural weighting schemes give 
similar super profile parameter values in 
the residual-scaled case, but values differ slightly for the non-residual scaled cubes.
\end{justify}
{\renewcommand{\arraystretch}{0.86}
\begin{deluxetable}{l c c c c c}
\centering
\tabletypesize{\scriptsize}
\tablewidth{0pt}
\tablecaption{The Sample galaxies \label{tab:sample}}
\tablehead{
\multicolumn{1}{c}{Galaxy}&\multicolumn{1}{c}{Weighting}
&\multicolumn{1}{c}{$B_{maj}$}&$B_{min}$ & Ch. width\\ 
& &(\arcsec)&(\arcsec)&$(\rm{km~ s^{-1}})$ \\
\multicolumn{1}{c}{1}&\multicolumn{1}{c}{2}&\multicolumn{1}{c}{3}& 4 &5 
}
\startdata
NGC 628 & NA & 11.88 & 9.30 & 2.6 \\
  & RO & 6.8 & 5.57 & \\
NGC 925 & NA & 5.94 & 5.71 & 2.6 \\
  & RO & 4.85 & 4.65 & \\
NGC 2366 & NA & 13.10 & 11.85 & 2.6\\ 
  & RO & 6.96 & 5.94 & \\
NGC 2403 & NA & 8.75 & 7.65 & 5.2\\ 
  & RO & 6.01 & 5.17 &\\  
Ho II & NA & 13.74 & 12.57 & 2.6\\ 
  & RO & 6.95 & 6.05 & \\
M81 dwA & NA & 15.87 & 14.23 & 1.3 \\
  & RO & 7.79 & 6.27 & \\
DDO 53 & NA & 11.75 & 9.53 & 2.6 \\
  & RO & 6.34 & 5.67 & \\
NGC 2903 & NA & 15.27 & 13.32 & 5.2 \\
  & RO & 8.66 & 6.43 & \\
Ho I & NA & 14.66 & 12.73 & 2.6 \\ 
  & RO & 7.78 & 6.03 & \\
NGC 2976 & NA & 7.41 & 6.42 & 5.2\\ 
  & RO & 5.25 & 4.88 & \\
NGC 3184 & NA & 7.51 & 6.93 & 2.6 \\
  & RO & 5.33 & 5.11 &\\ 
NGC 3198 & NA & 13.01 & 11.56 & 5.2\\ 
  & RO & 7.64 & 5.62 & \\
IC 2574 & NA & 12.81 & 11.90 & 2.6\\ 
  & RO & 5.93 & 5.48 & \\
NGC 3351 & NA & 9.94 & 7.15 & 5.2 \\
  & RO & 6.26 & 5.20 &\\
NGC 3621 & NA & 15.95 & 10.24 & 5.2\\
  & RO & 10.50 & 5.68 &\\ 
NGC 4214 & NA & 14.69 & 13.87 & 1.3\\ 
  & RO & 7.41 & 6.35 & \\
NGC 4736 & NA & 10.22 & 9.07 & 5.2\\ \
  & RO & 5.96 & 5.55 &\\ 
DDO 154 & NA & 14.09 & 12.62 & 2.6 \\
  & RO & 7.94 & 6.27 & \\
NGC 5055 & NA & 10.06 & 8.66 & 5.2 \\
  & RO & 5.78 & 5.26 & \\
NGC 5236 & NA & 15.16 & 11.44 & 2.6\\ 
  & RO & 10.40 & 5.60 & \\
NGC 6946 & NA & 6.04 & 5.61 & 2.6\\ 
  & RO & 4.93 & 4.51 &\\
NGC 7793 & NA & 15.60 & 10.85 & 2.6 \\
  & RO & 10.37 & 5.39 &\\  
\enddata
\tablecomments{Column 1: Name of galaxy; Column 2: NA: Natural weighting; RO: Robust weighting; Column 3/4: 
major and minor axis of synthesized beam in arcsec; Column 5: Channel width.} 
\end{deluxetable}
}
\vspace{-1cm}
\subsection{Non-residual vs residual-scaled cubes}
\begin{figure*}
\centering
    \begin{tabular}{c c c}
    \includegraphics[scale=.27]{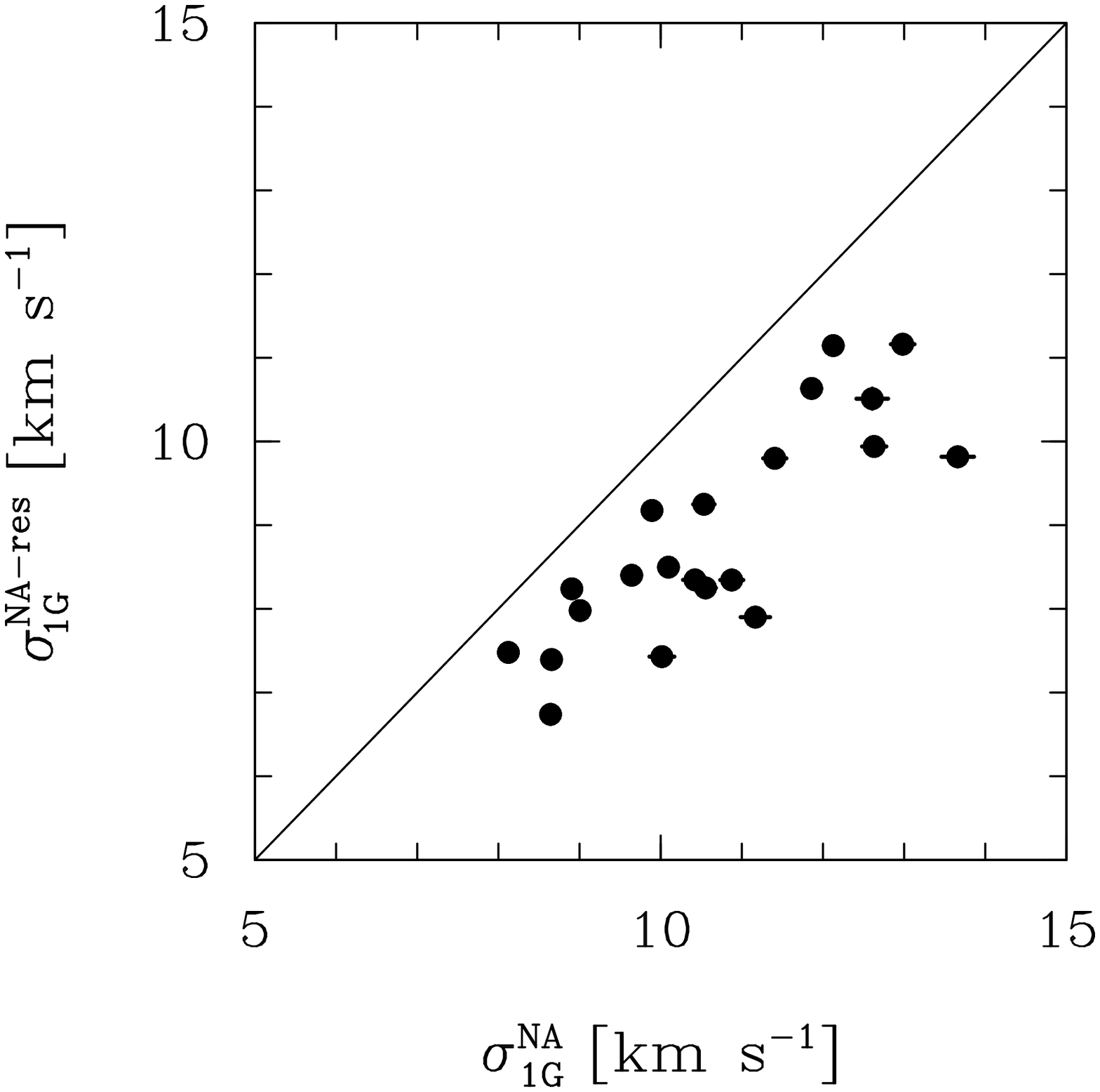}& 
  
     \includegraphics[scale=.27]{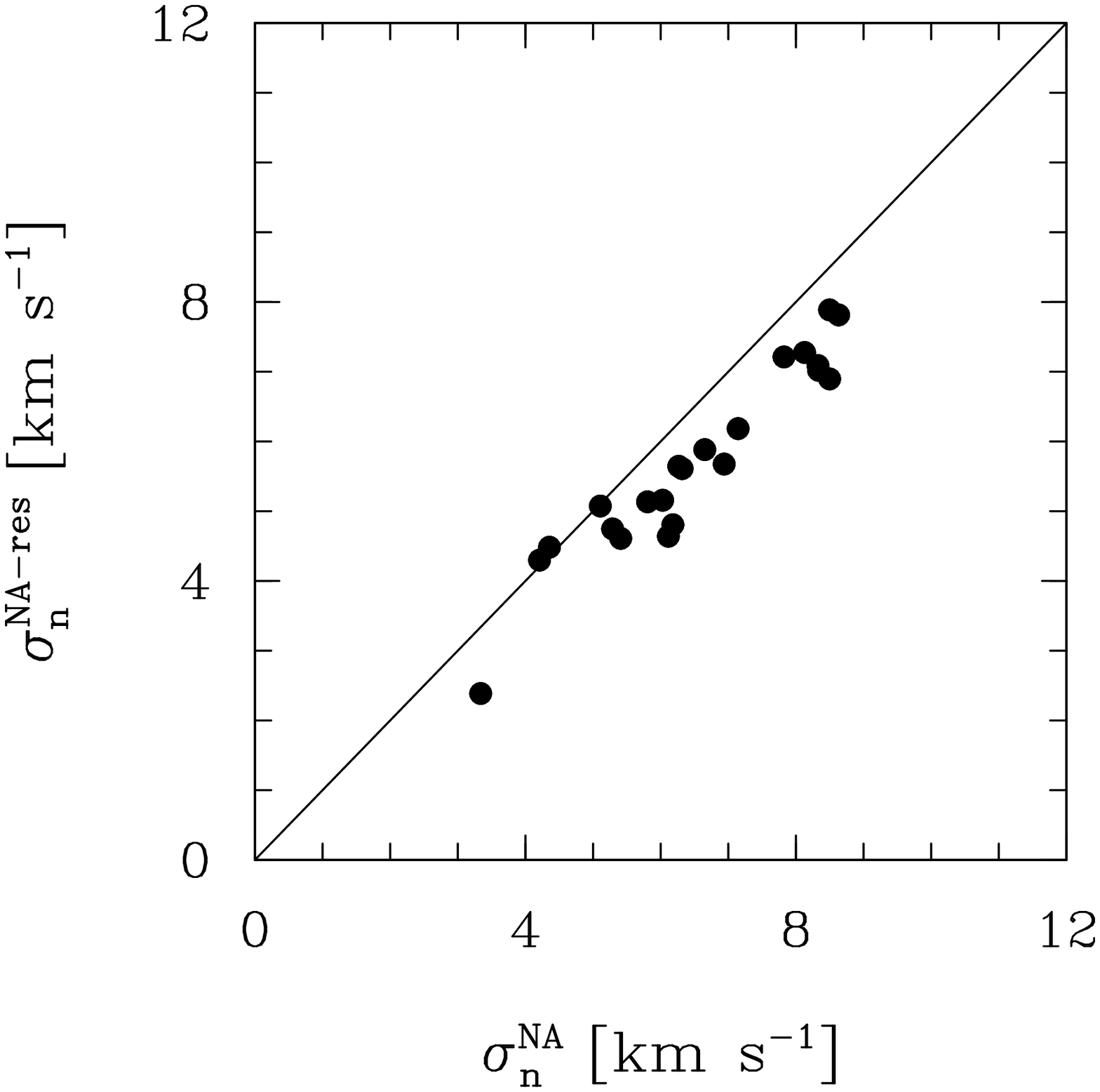}&
  
     \includegraphics[scale=.27]{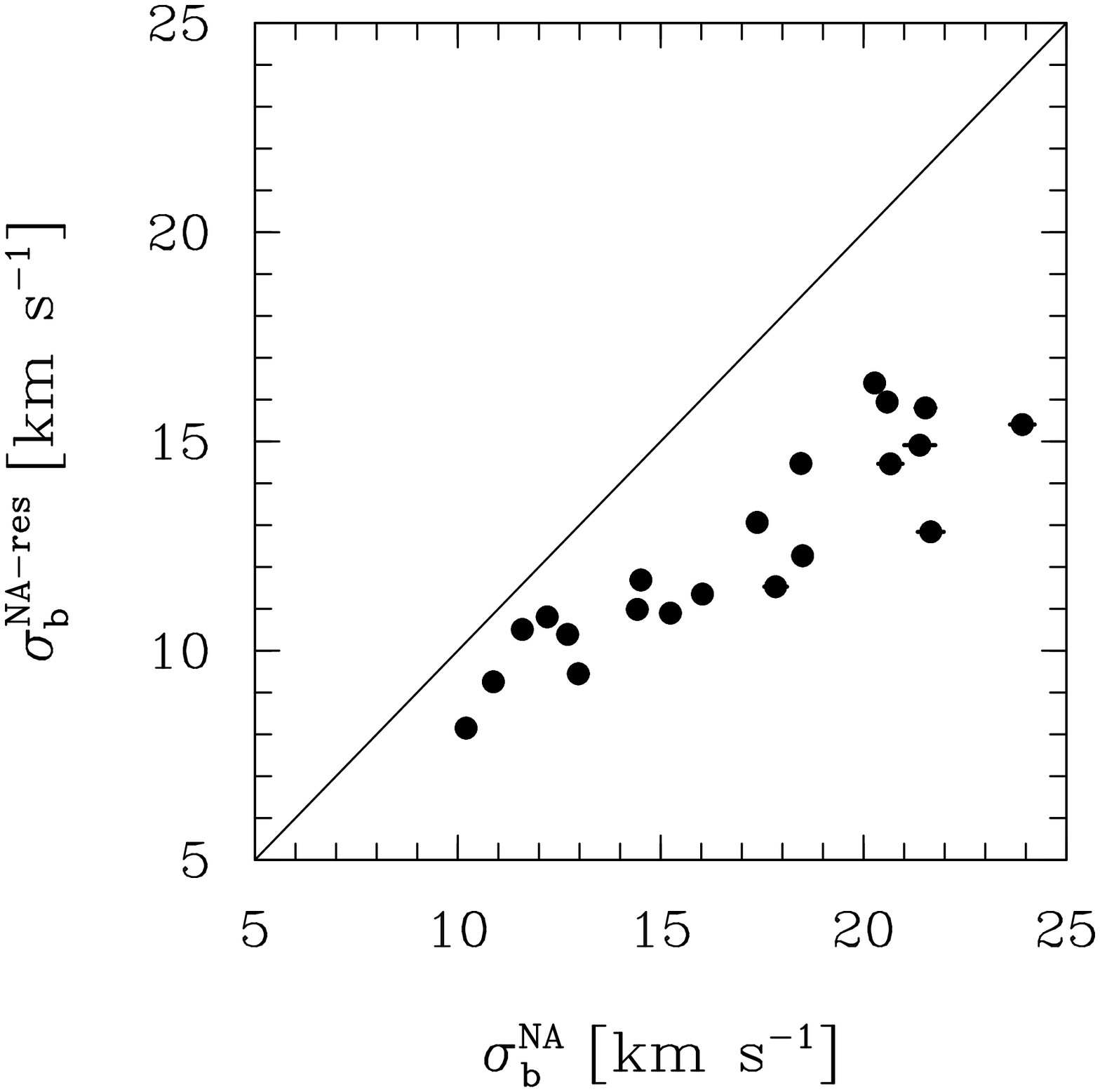}\\[.1in]
     \includegraphics[scale=.27]{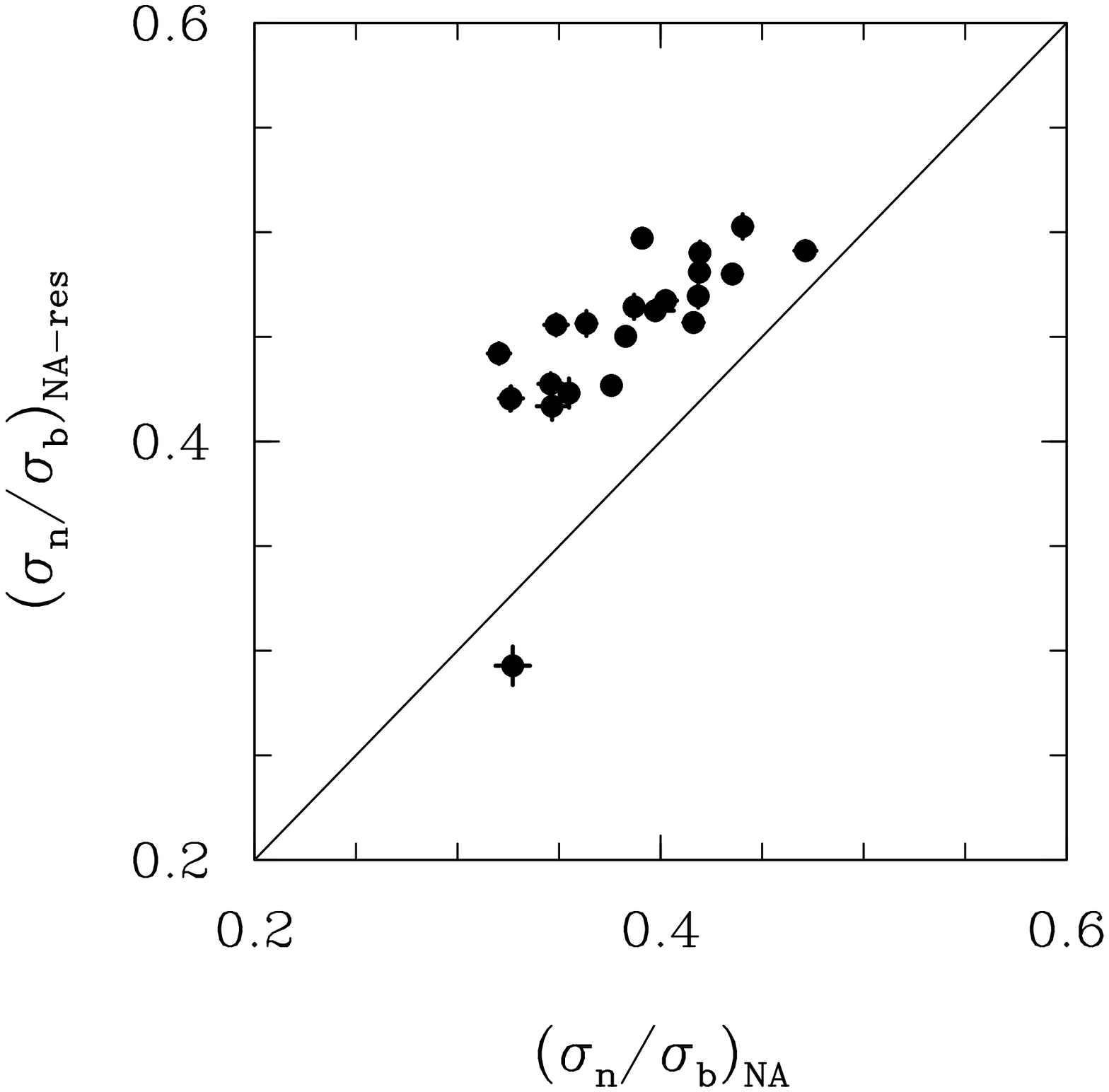}&
 
     \includegraphics[scale=.27]{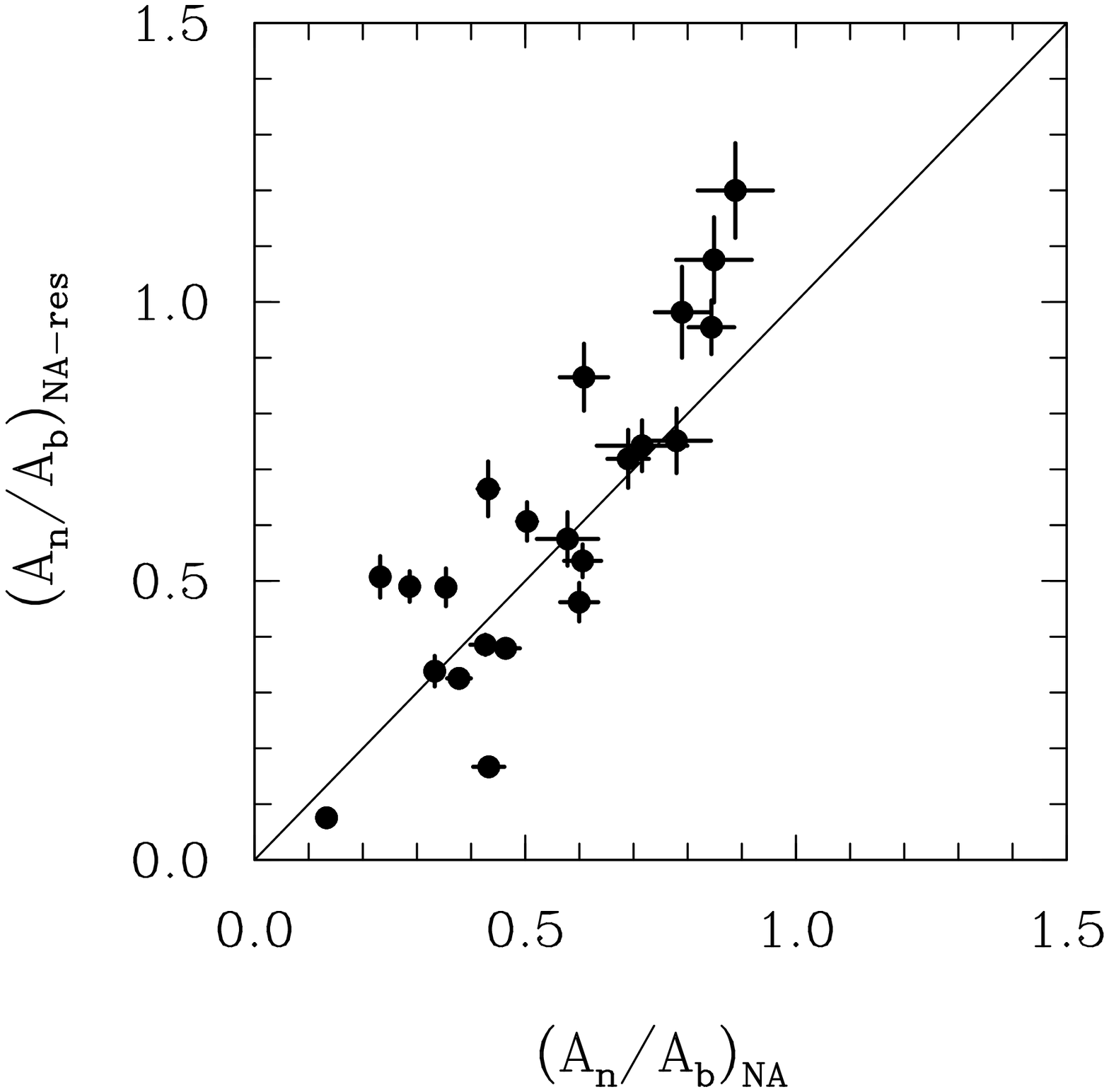}
    \end{tabular}
    \caption{Super profile parameters from natural residual-scaled 
    (NA-res) and natural non-residual scaled (NA) data cubes.} 
    \label{fig:comp_na_res_nonres}
\end{figure*}

In this section, we compare super profile parameters derived from non-residual scaled cubes with those from residual-scaled cubes. In Figure~\ref{fig:comp_na_res_nonres}, we compare super profile parameters of the Natural non-residual scaled 
cubes (NA) with those of the Natural residual-scaled cubes (NA-res). The NA cubes give mean single Gaussian and narrow component velocity dispersions that are respectively higher by $\sim$~18\% and $\sim$~12\% than the NA-res cubes. For the broad component dispersions, the NA cubes give a mean value that is $\sim$~27\% higher than that of the NA-res cubes.  As a result, the NA cubes  give smaller $\sigma_{n}/\sigma_{b}$ values than the NA-res ones. 
The $A_{n}/A_{b}$ values from the NA and NA-res cubes are similar. 

\begin{figure*}
\centering
    \begin{tabular}{c c c c}
    \includegraphics[scale=.27]{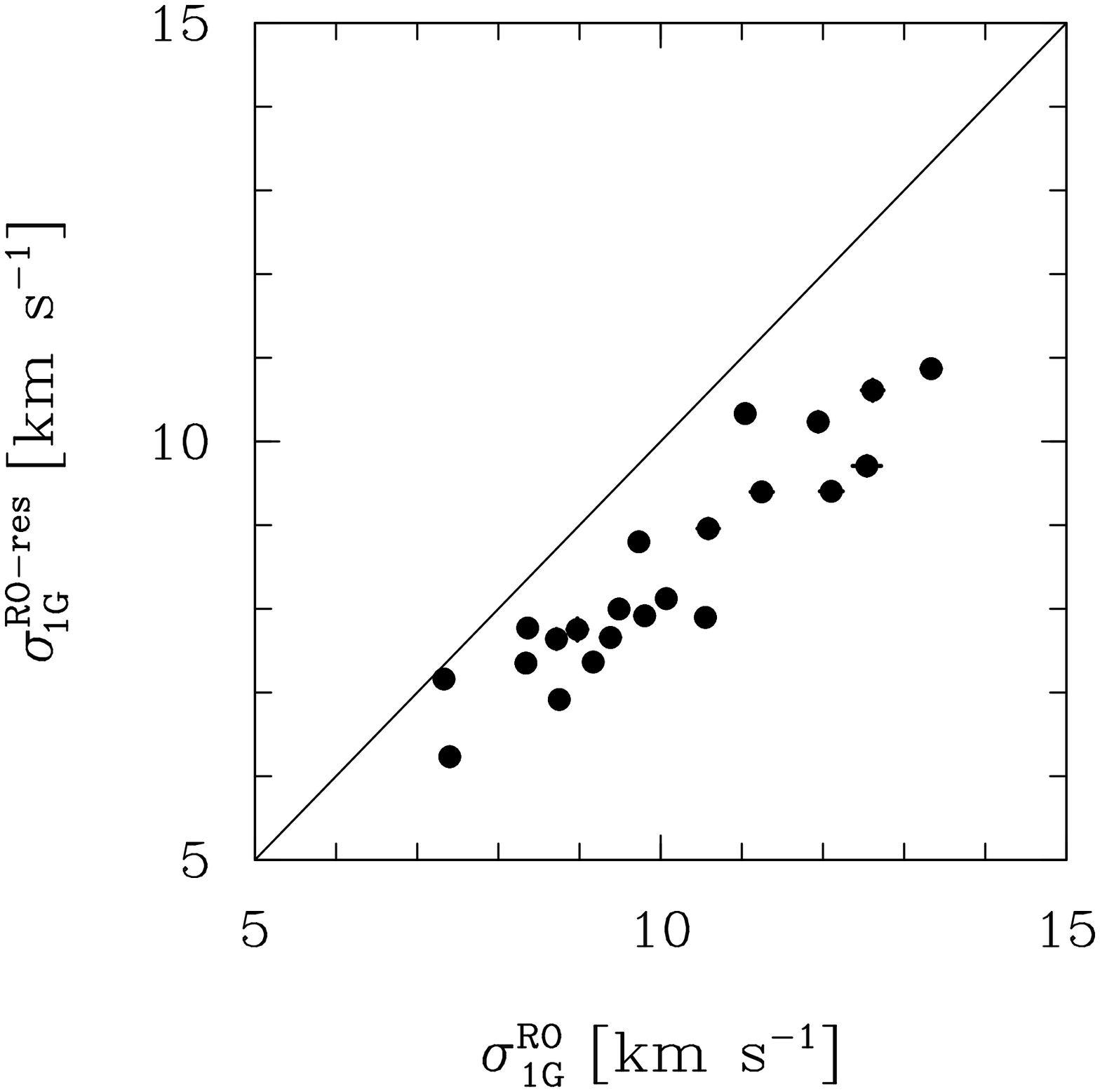}& 
   
     \includegraphics[scale=.27]{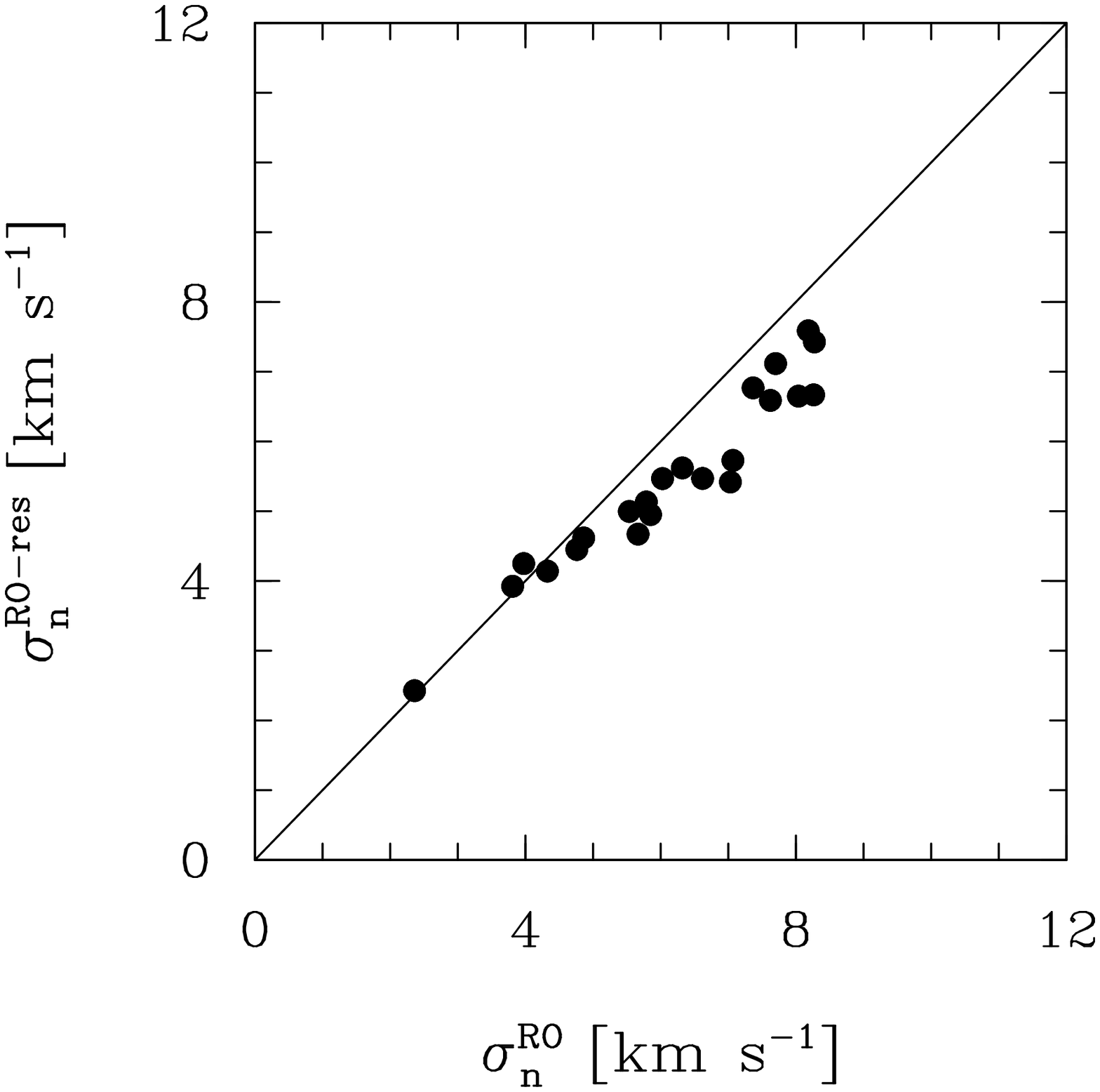}&

     \includegraphics[scale=.27]{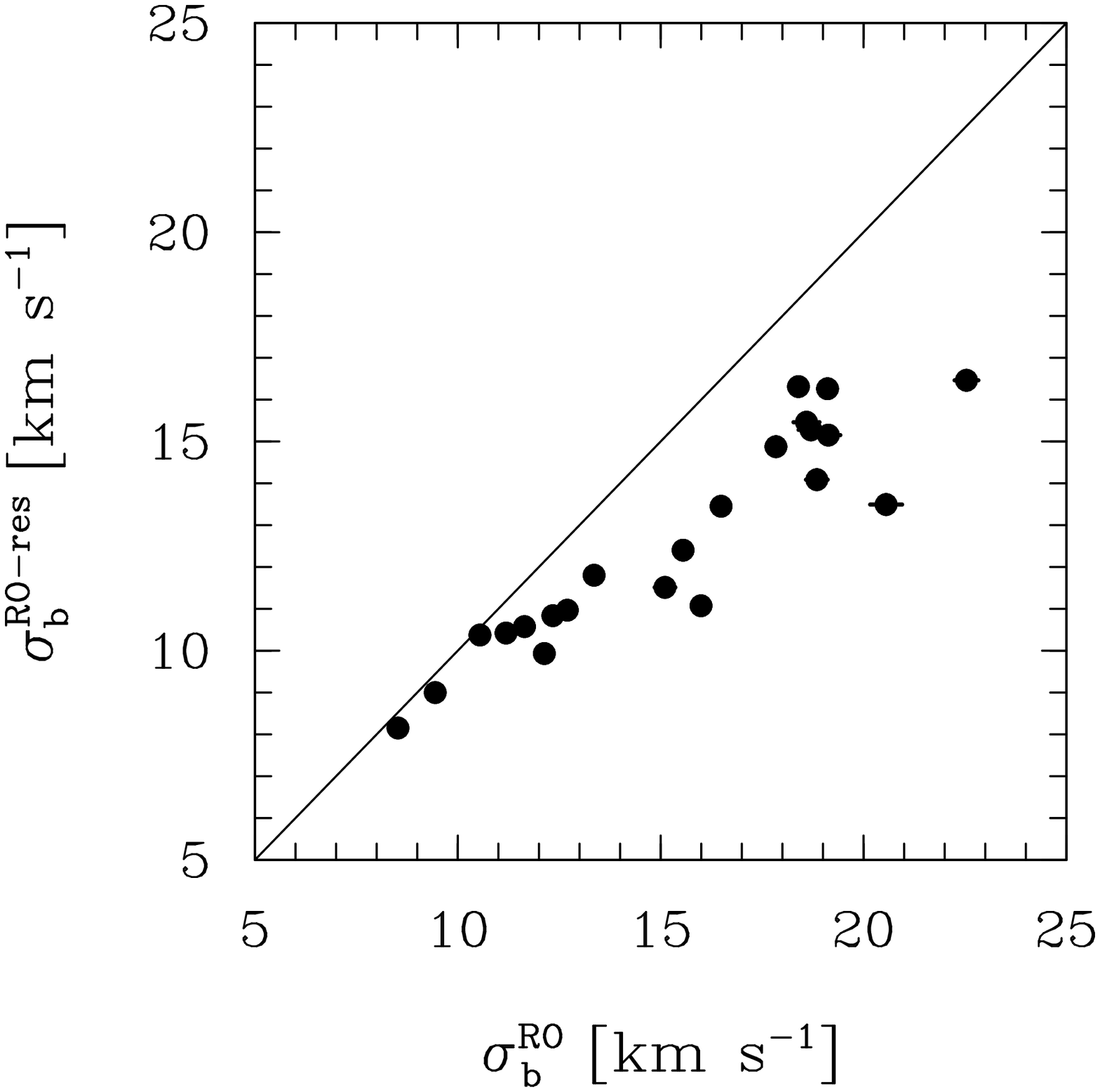}\\[.1in]
     \includegraphics[scale=.27]{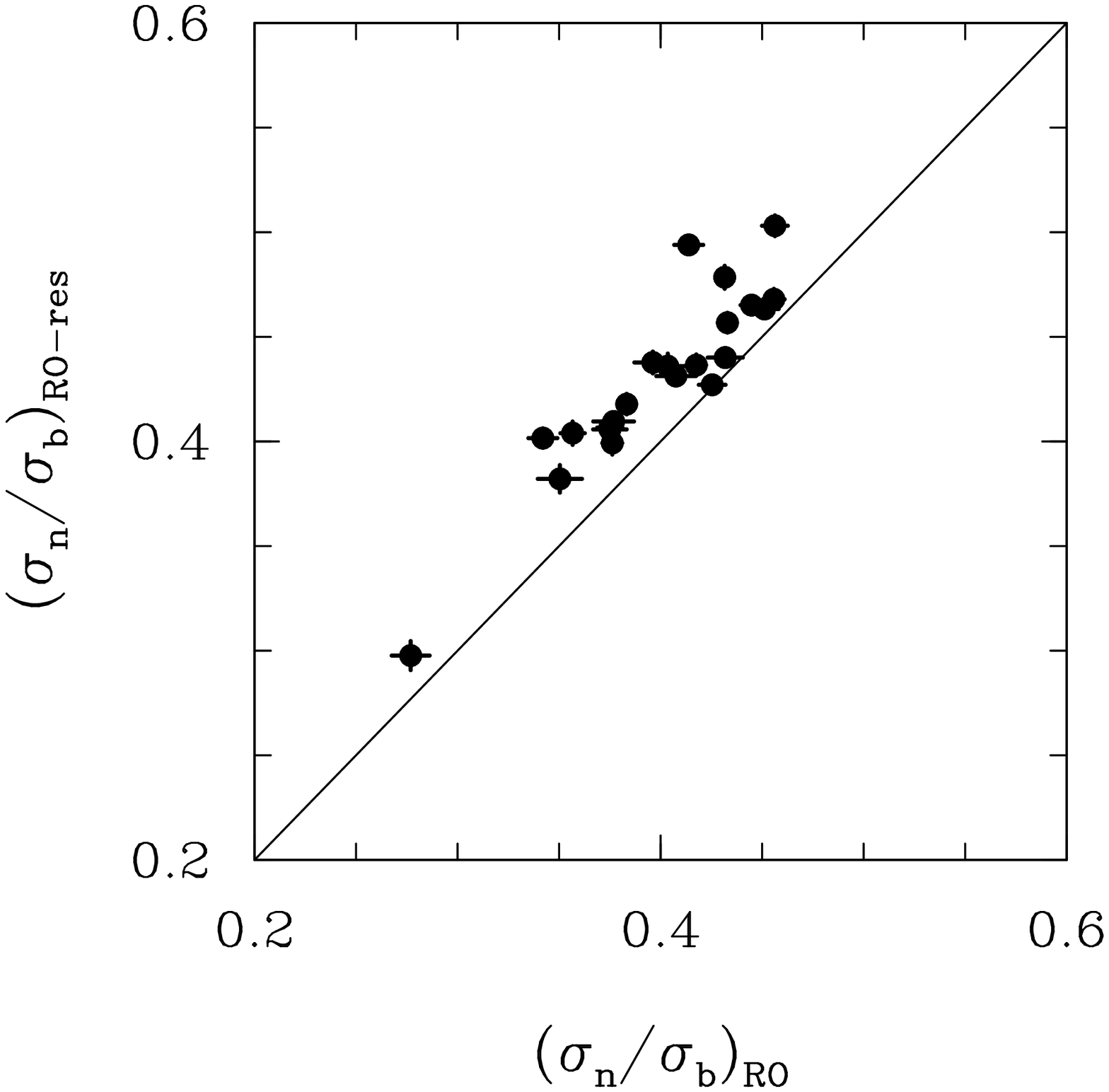}&

     \includegraphics[scale=.27]{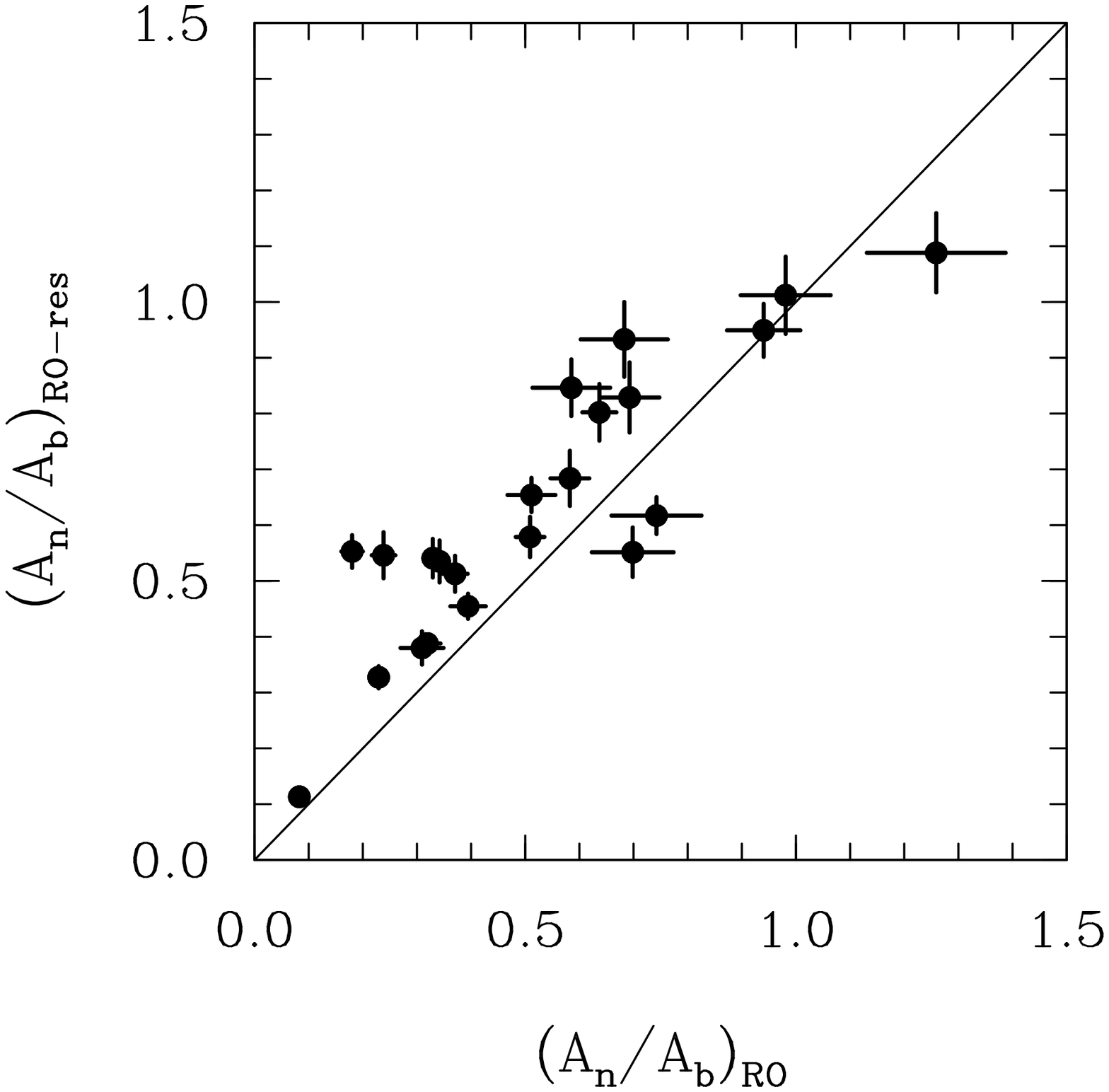}
    \end{tabular}
    \caption{Comparison of the super profile parameters from robust non-residual scaled (RO) 
    and robust residual-scaled (RO-res) cubes.}  
    \label{fig:comp_ro_res_nonres}
\end{figure*}

In Figure~\ref{fig:comp_ro_res_nonres}, we compare parameters from Robust non-residual scaled 
cubes (RO) and Robust residual-scaled cubes (RO-res). Here the $\sigma_{n}$ from the RO 
and the RO-res cubes are similar. The mean $\sigma_{1G}$ and $\sigma_{b}$ values from the RO are, 
on average, $\sim$~16\% higher than that of the RO-res, which are much lower than 
the difference found for the Natural-weighted cubes. 

In conclusion, the broad component is more sensitive to residual-scaling effects than 
the narrow and the single Gaussian components. 
Similarly, the effects of residual-scaling are more pronounced for the Natural weighted than for the 
Robust weighted cubes. Natural non-residual scaled cubes give broad component 
dispersion values that are on average 10\% higher than that of Robust non-residual scaled cubes. 
Super profile parameters from residual-scaled cubes are similar despite the weighting scheme adopted.  
In general, the broad component dispersion values from non-residual scaled 
cubes are higher than those from residual-scaled cubes. 
\subsection{Blanked vs non-blanked cubes}
\begin{figure*}
\centering
    \begin{tabular}{c c c}
    \includegraphics[scale=.27]{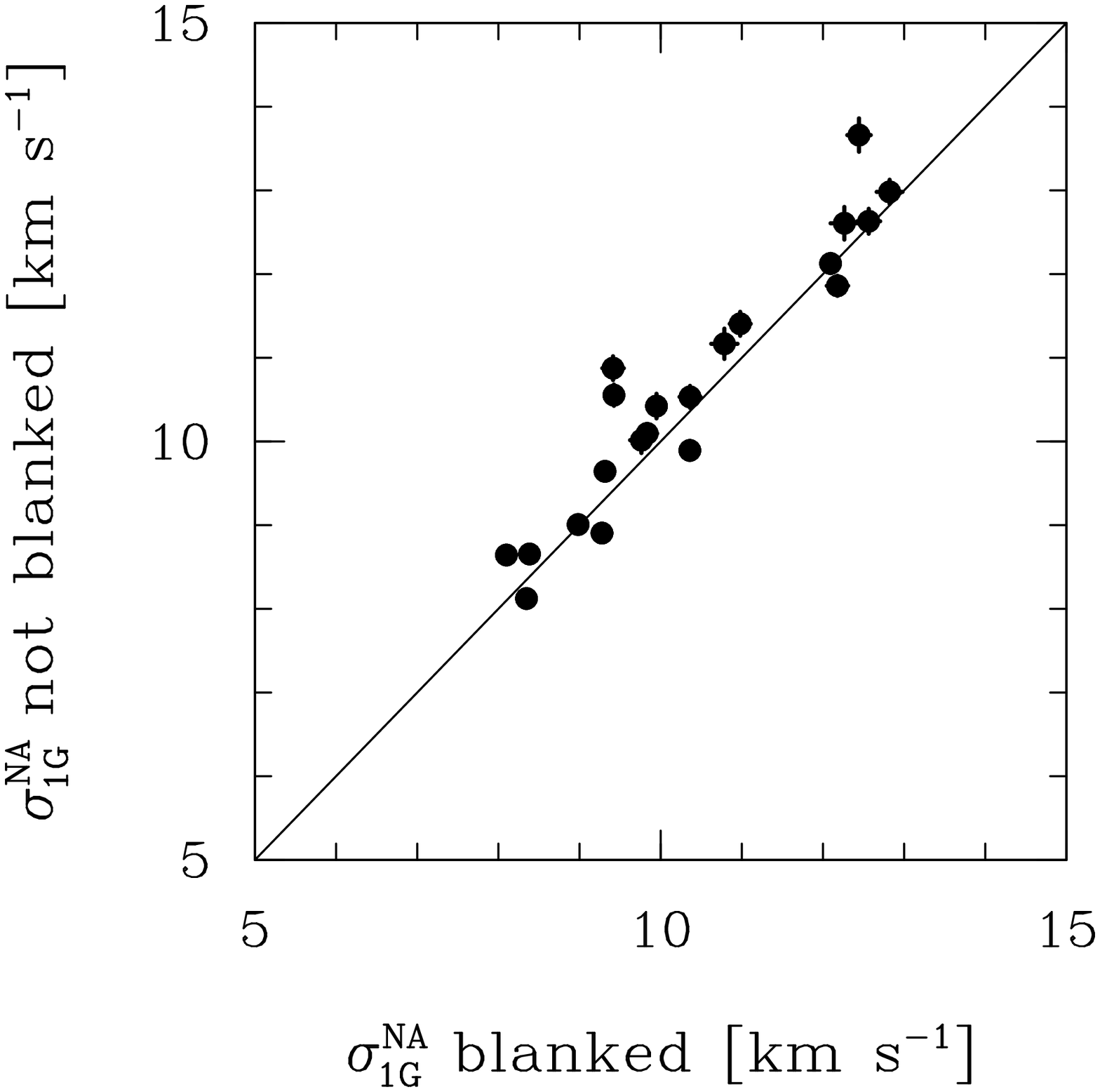}& 

     \includegraphics[scale=.27]{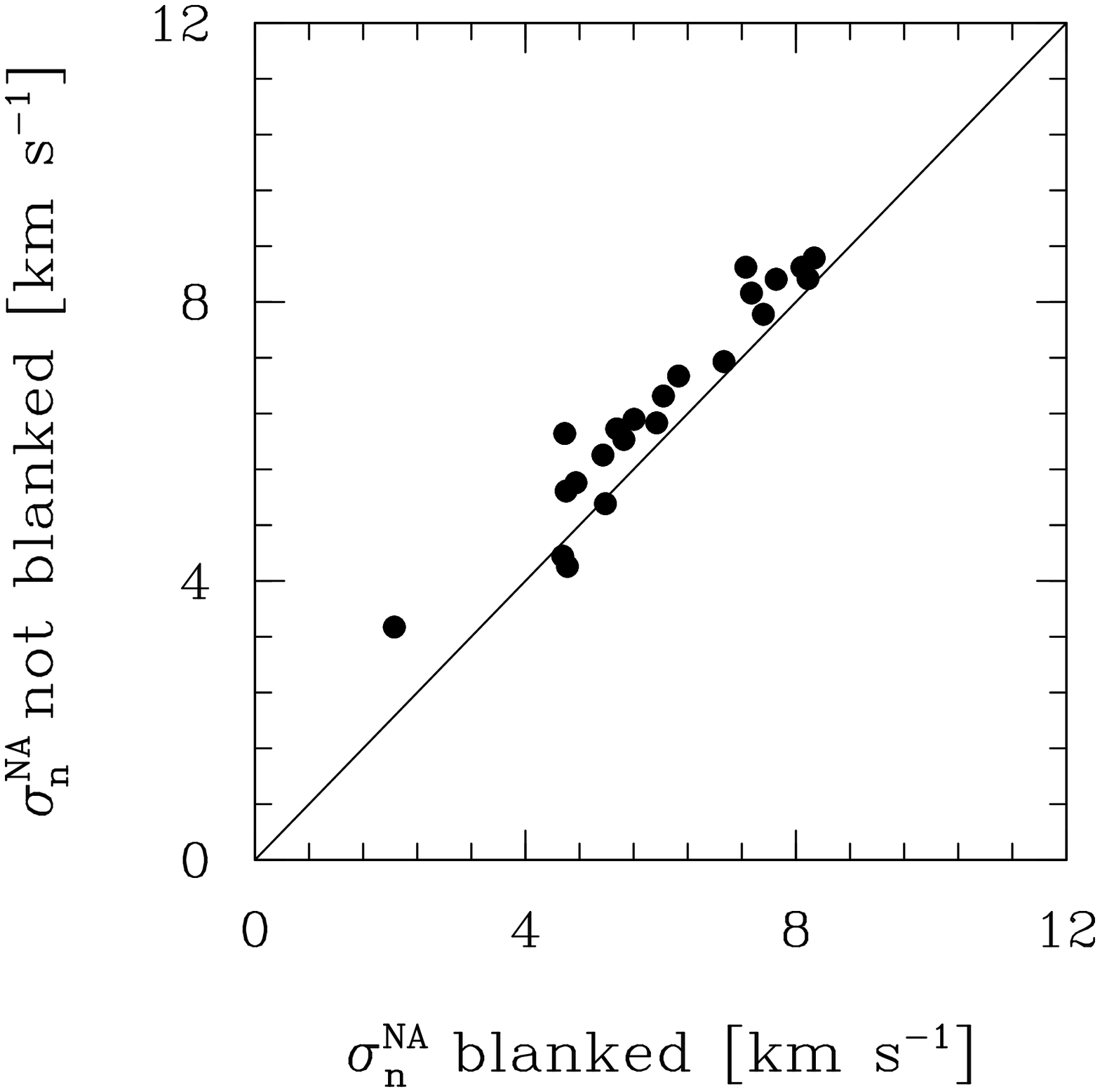}&

     \includegraphics[scale=.27]{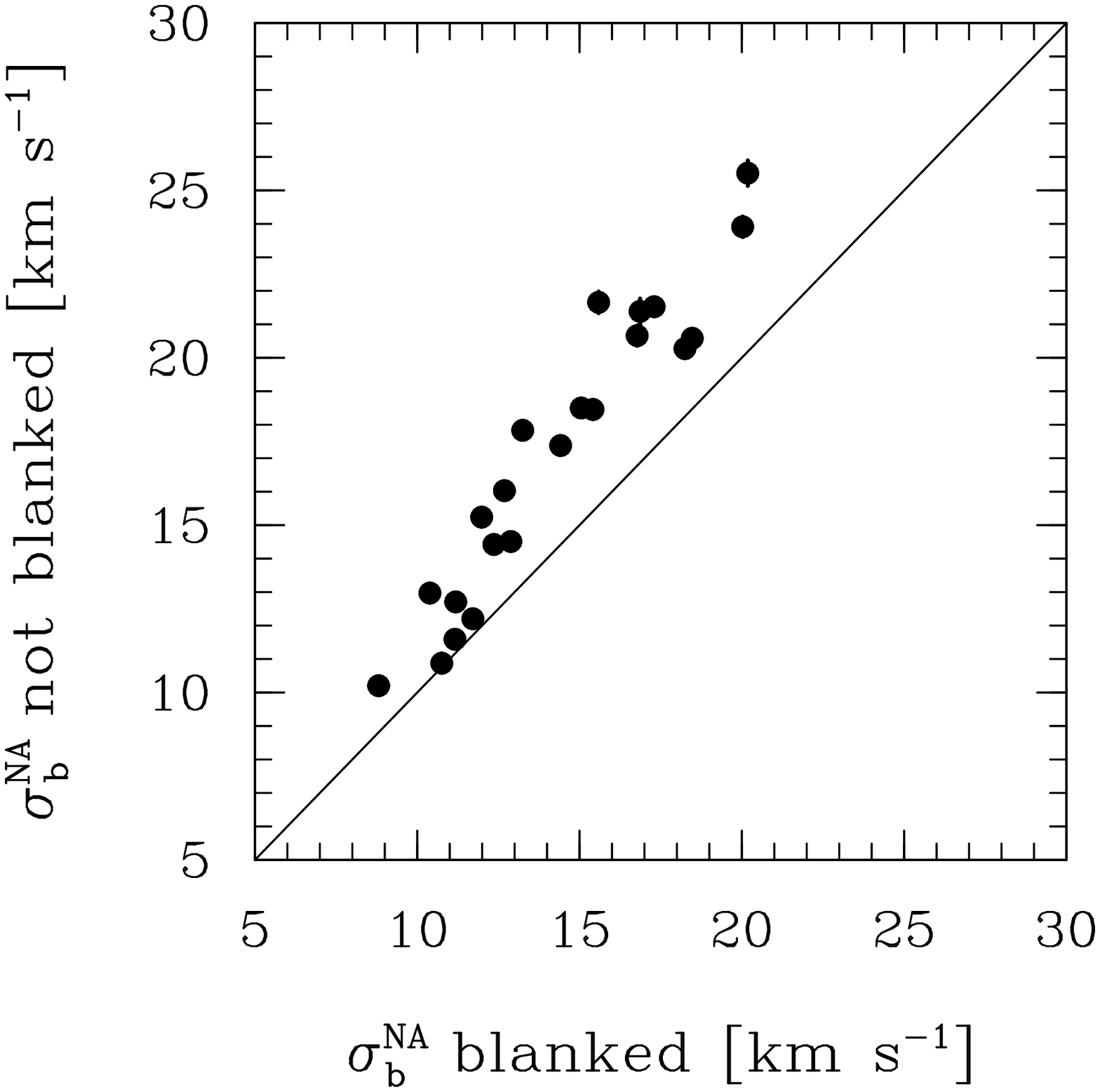}\\

     \includegraphics[scale=.27]{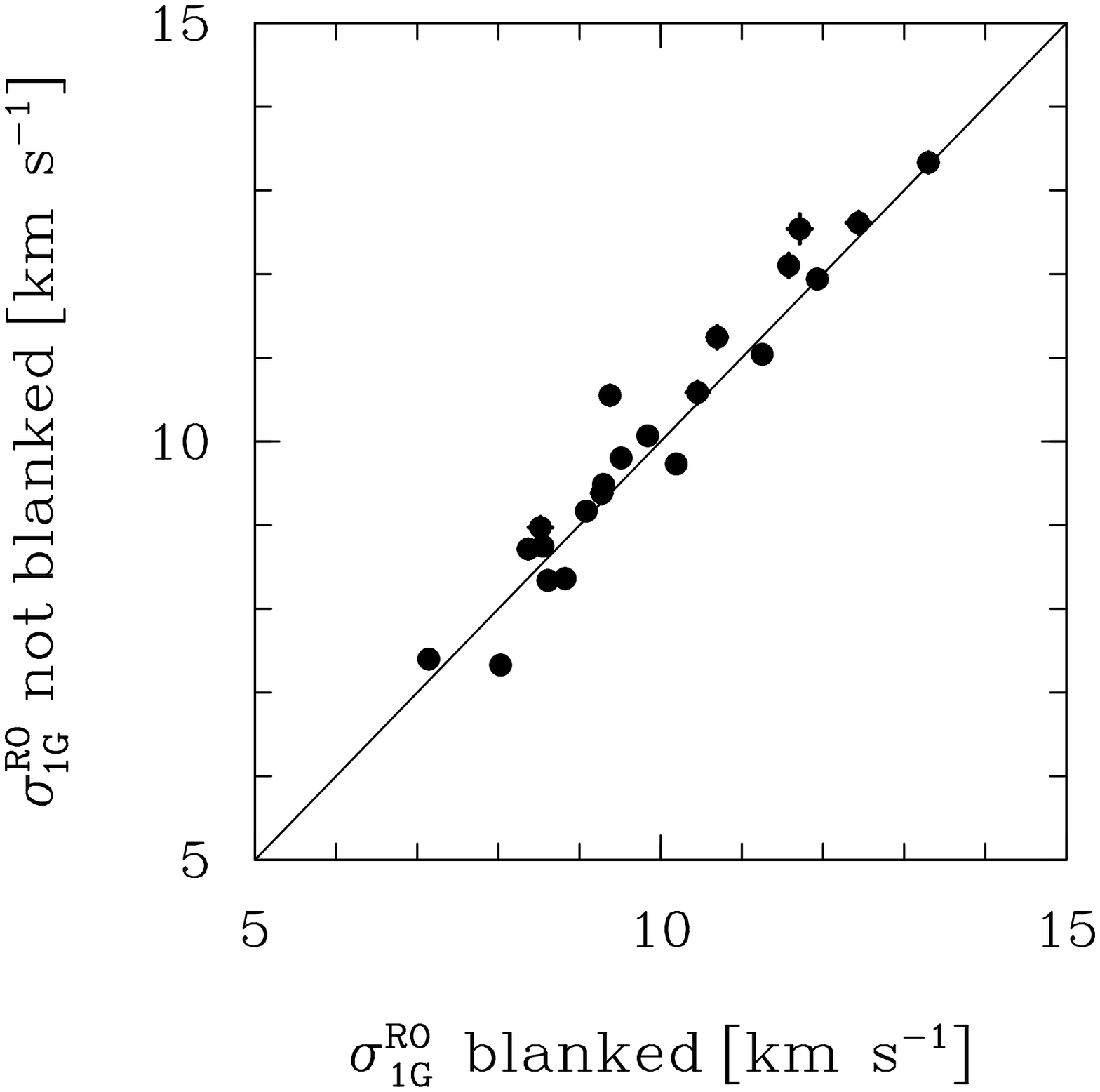}&
     \includegraphics[scale=.27]{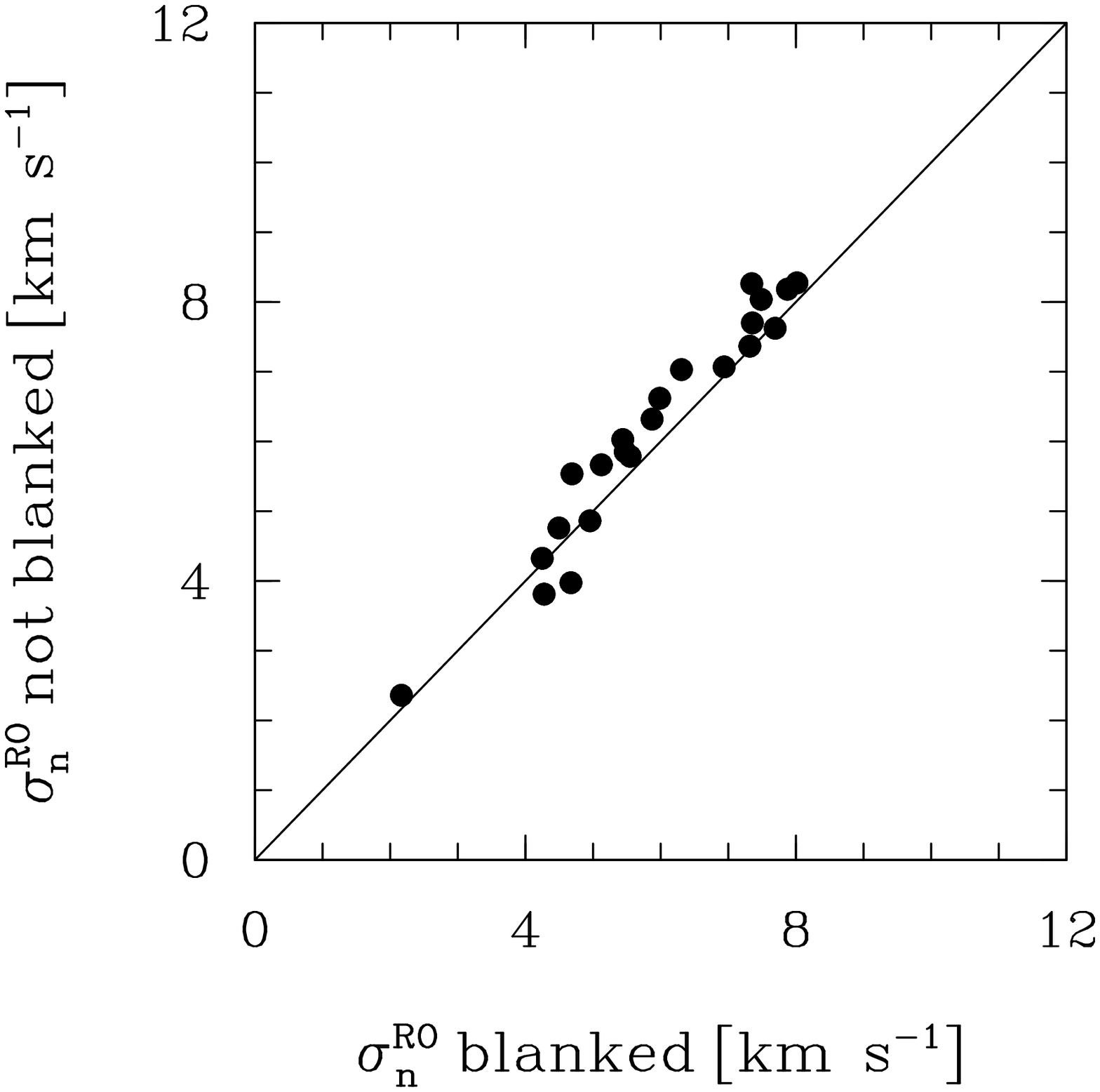}&
     \includegraphics[scale=.27]{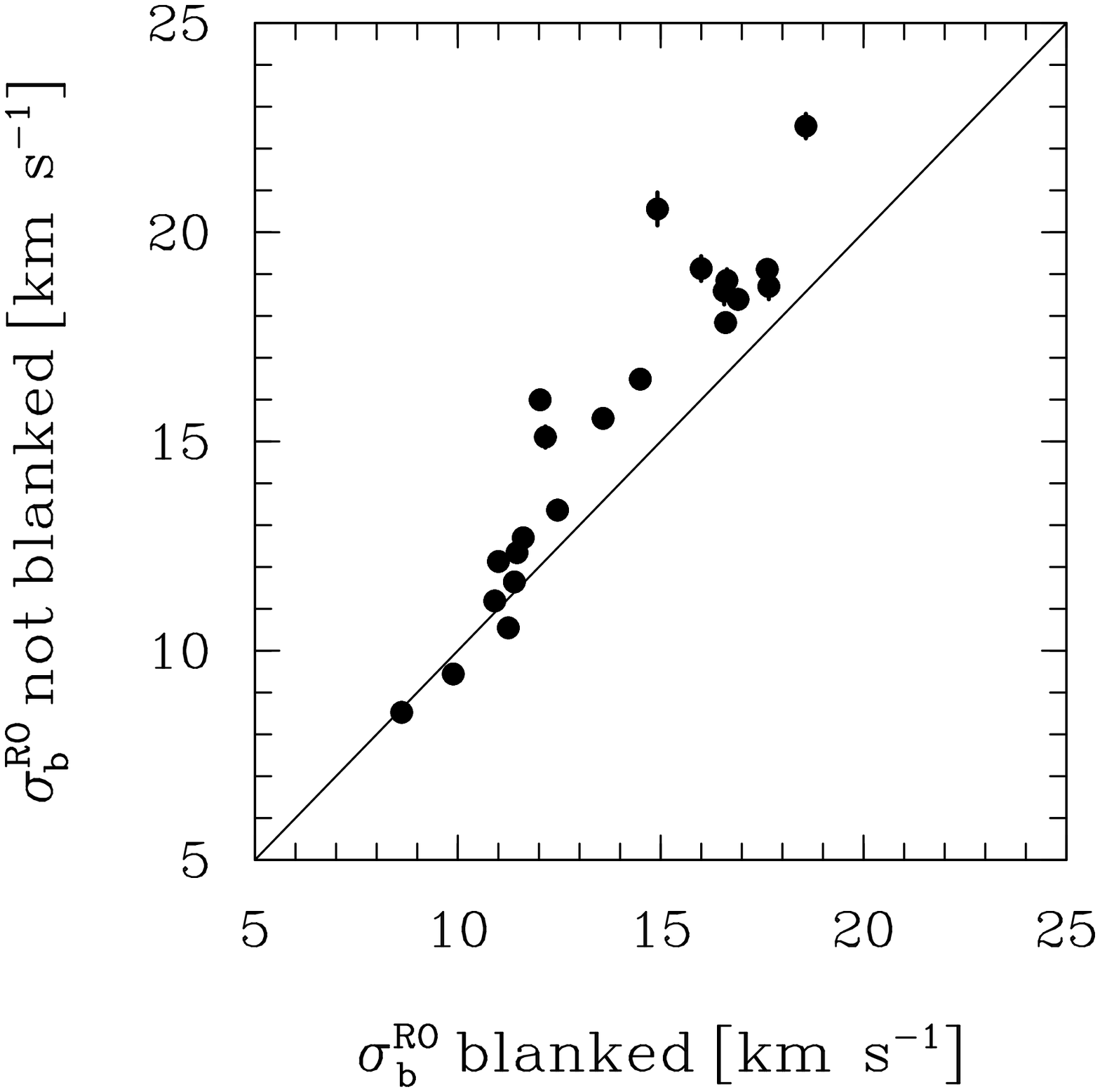}
    \end{tabular}
    \caption{Blanked non-residual scaled cubes vs non-blanked non-residual scaled cubes.} 
    \label{fig:comp_blanonres_nonblanonres}
\end{figure*}

\begin{figure*}
\centering
    \begin{tabular}{c c c}
    \includegraphics[scale=.27]{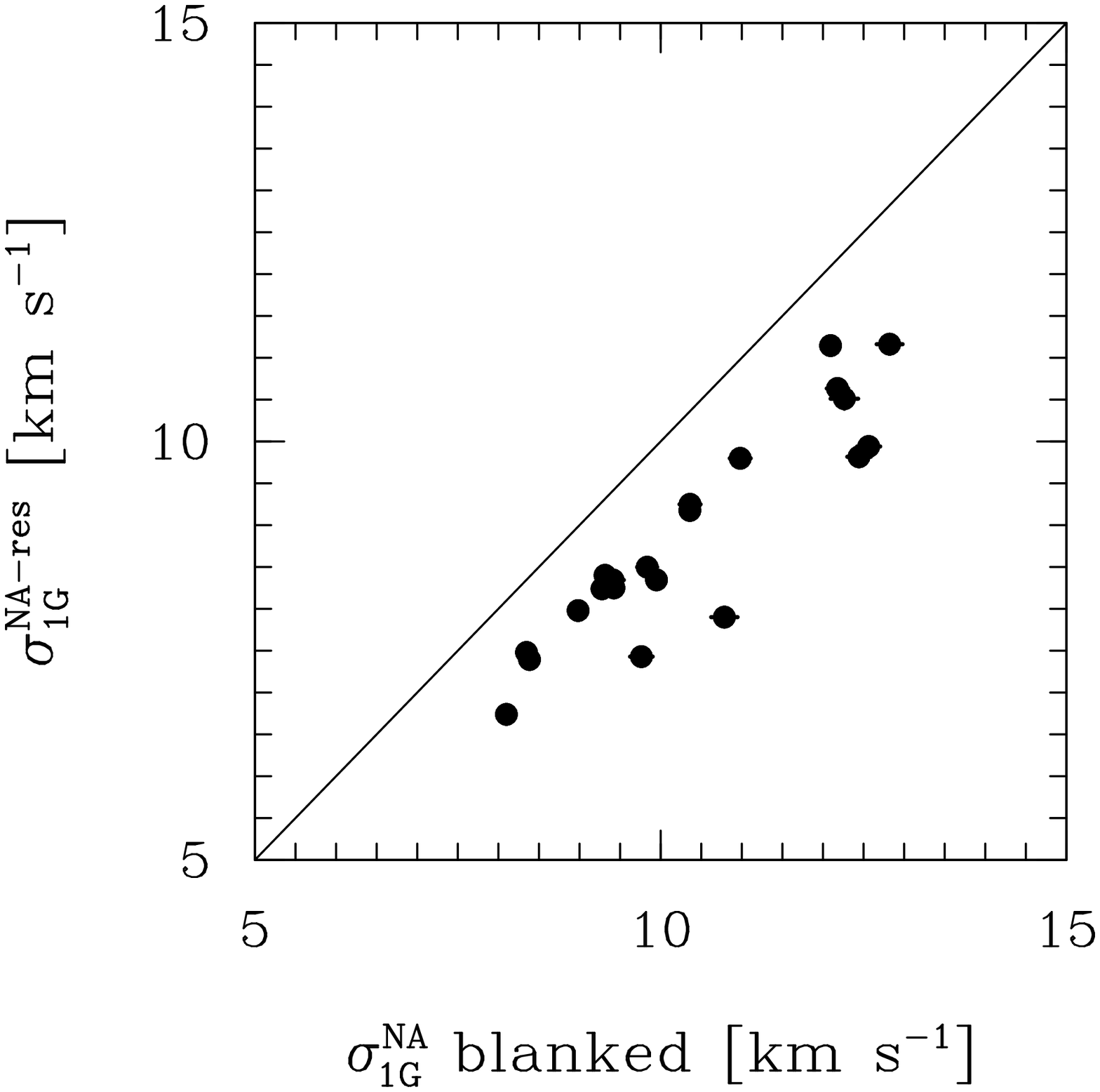}& 
     \includegraphics[scale=.27]{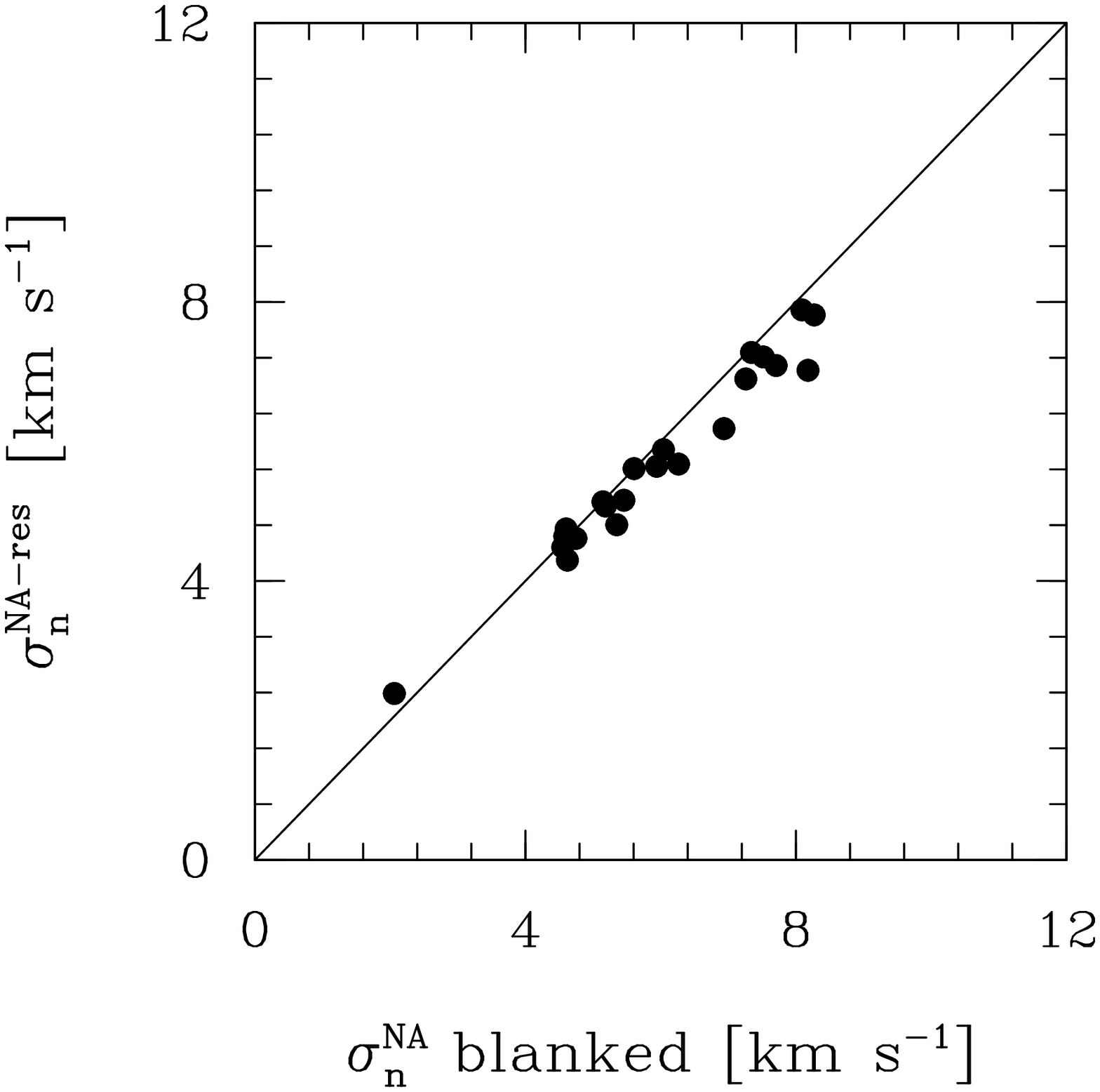}&
     \includegraphics[scale=.27]{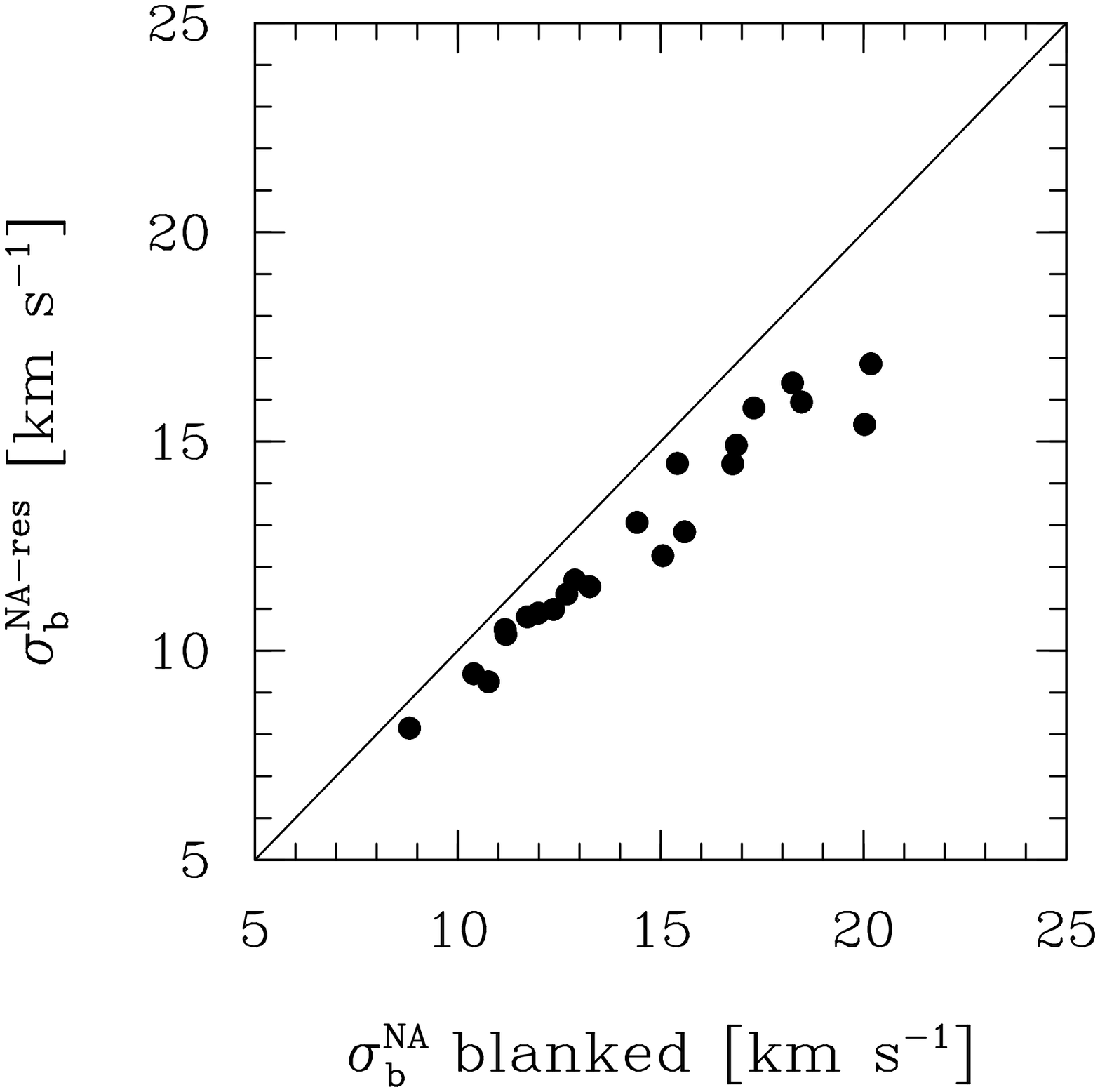}\\
     \includegraphics[scale=.27]{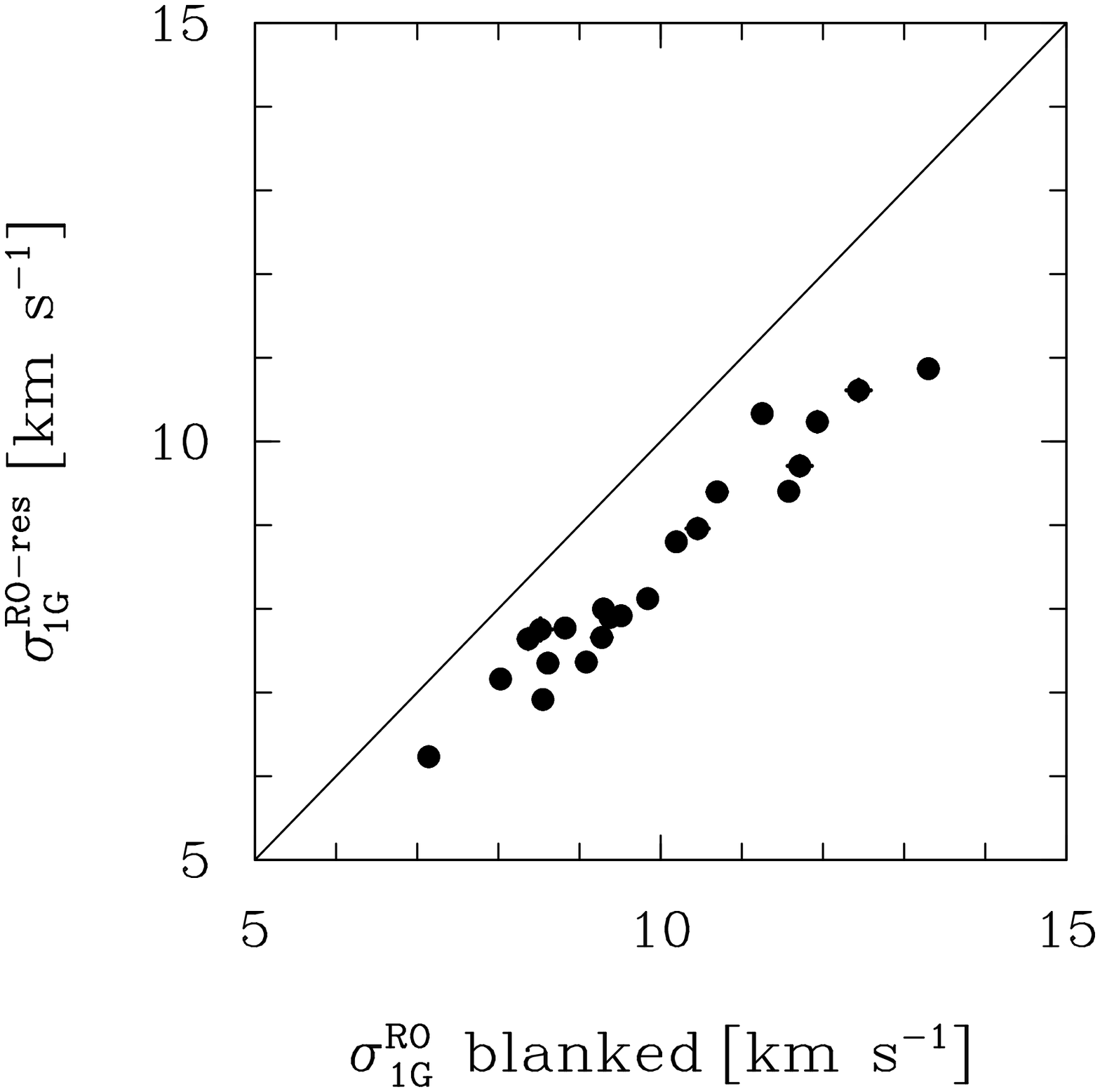}&
     \includegraphics[scale=.27]{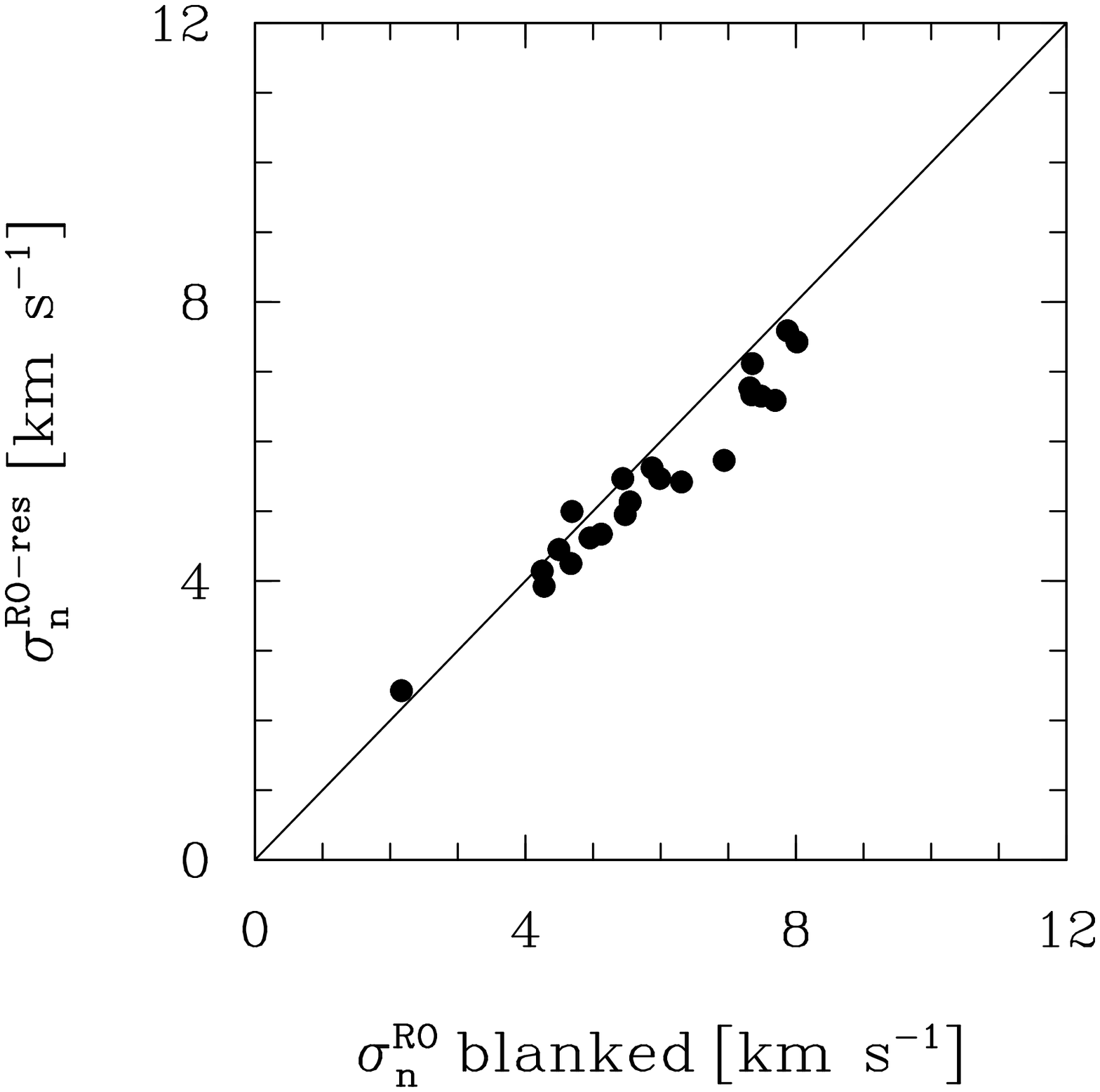}&
     \includegraphics[scale=.27]{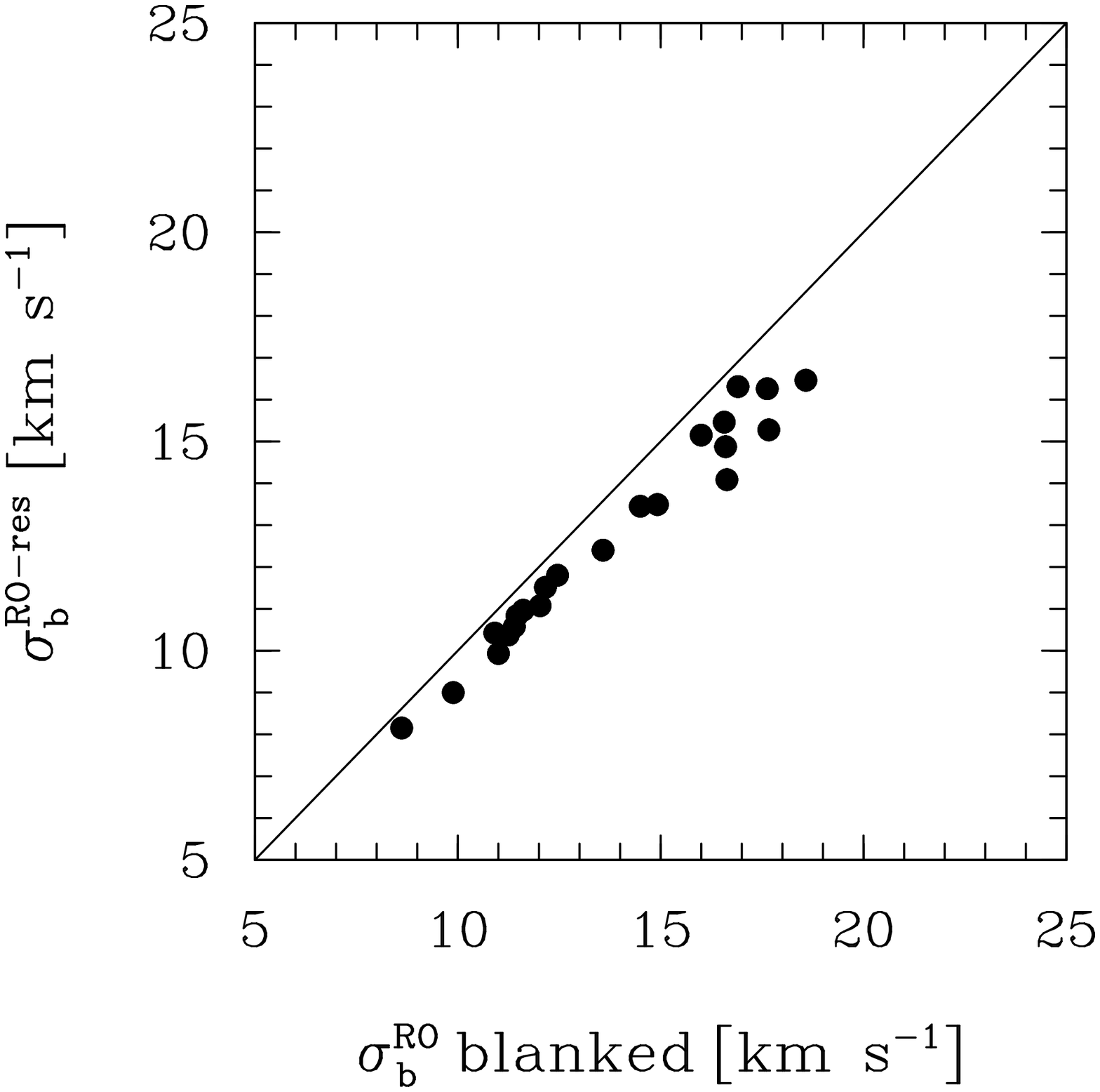}
    \end{tabular}
    \caption{Blanked non-residual scaled cubes vs blanked residual-scaled cubes.} 
    \label{fig:comp_blanonres_blares}
\end{figure*}

In the previous section, we have shown that residual-scaled cubes tend to give lower velocity dispersion values 
than non-residual scaled cubes. Note, however, that as 
already mentioned in Section~\ref{sec:res_scale} residual-scaled cubes are also 
blanked \citep{walter08}. So how much of the difference is then caused by blanking and how much is due to 
the residual scaling? To investigate this, we blank the non-residual 
scaled cubes using the same blanking mask as the residual-scaled cubes. 
We do this for both the Robust and the Natural-weighted data cubes, then derive 
and fit super profiles from these blanked non-residual scaled cubes. 
In Figure~\ref{fig:comp_blanonres_nonblanonres}, we compare the velocity dispersions 
from the blanked non-residual scaled cubes with those 
from the non-blanked non-residual scaled cubes. 
In Figure~\ref{fig:comp_blanonres_blares}, we compare the velocity 
dispersions from the blanked non-residual scaled cubes with those from the 
(already blanked) residual-scaled cubes. We conclude from these two figures that blanking tends 
to affect only the broad component. This can be understood by the fact that the 
broad component mostly represents faint emission, which manifests as wings in 
the super profiles. This low-level emission consists mostly of uncleaned residual emission, which leads 
to an overestimate of the broad component velocity dispersion when not 
blanked or residual-scaled. Thus, the difference in broad component velocity dispersion between 
the residual-scaled cubes and the non-residual residual scaled cubes presented in the 
previous section is partly due to the effects of blanking. For the single Gaussian 
velocity dispersion, which is most commonly used in the literature, 
blanking the data cubes does not affect the velocity dispersion, whereas 
residual-scaling the cubes does. 
  
 \subsection{Effects of CLEANing level on H\,{\sc i} velocity dispersions 
 for residual-scaled and non-residual scaled cubes}
 In the previous sections, we have analyzed the effects of weighting and 
 residual scaling on the global measurement of the H\,{\sc i} velocity dispersion 
(i.e., from the entire disk of the galaxies) from data cubes cleaned to 2.5 
times the rms noise. However, accurate measurement of the H\,{\sc i} velocity dispersion 
as a function of radius is also important \citep[e.g., 
for the analysis of disk (in)stability and/or the studies of the shape of the dark matter halo,][]{petricrupen}. 
Here we explore the effects of the choice of CLEAN depth on  
the measurement of H\,{\sc i} velocity dispersion as a function of radius.

\begin{figure*}
\centering
    \begin{tabular}{r}
    \includegraphics[scale=.2]{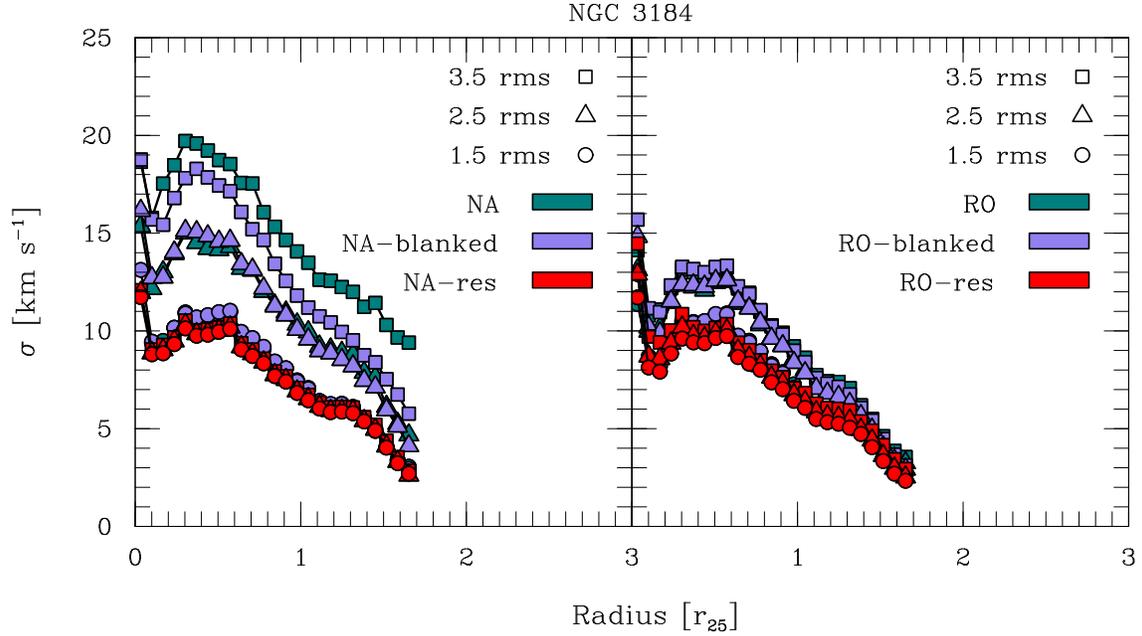}
    \end{tabular}
    \caption{Single Gaussian radial velocity dispersion profiles of NGC 3184 using different weighting schemes 
    and cleaning depths. Different symbols represent different cleaning levels. 
    \textit{Square symbols}: 3.5 times rms; \textit{triangle symbols}: 
    2.5 times rms \textit{circle symbols}: 
    1.5 times rms. \textit{Green}: non-residual scaled cubes (no blanking applied); \textit{purple}: 
    non-residual scaled and blanked cubes; \textit{red}: residual-scaled cubes. 
    \textit{Left panel}: Natural-weighted data cubes; \textit{right panel}: Robust-weighted data cubes.}
    \label{fig:clean}
\end{figure*}

We use the dirty image of NGC 3184, a face-on spiral galaxy, and clean it down to 3.5, 2.5 and 1.5 
times the rms noise using Natural and Robust weightings. The 2.5 times rms cube is identical 
to the standard THINGS cube. We also create blanked 
non-residual scaled and (blanked) residual-scaled versions of these cubes. 
We divide the image 
of NGC 3184 into series of elliptical annuli with 0.1 r$_{25}$ width 
\citep[we adopt r$_{25}$ = 222\arcsec,][]{walter08}, 
with identical orientation parameters (inclination, position angle, center position) 
as taken from \citet{walter08}. We stack individual profiles within each of these annuli 
and fit the resulting super profiles of the (3.5, 2.5, 1.5) times rms cubes with a single 
Gaussian function to derive the radial velocity dispersion profiles shown in Figure~\ref{fig:clean}. 
We show the shapes of the super profiles at a radius of 0.5 r$_{25}$ in Figure~\ref{fig:clean_prof}.  
We choose NGC 3184 because of its face-on orientation so that projection and beam smearing 
effects are minimal. For the Natural non-residual scaled cubes, there is a noticeable difference between 
the velocity dispersion derived from the (3.5, 2.5, 1.5) times rms cubes. 
This difference is smaller for the Robust non-residual scaled cubes. 
For residual-scaled data cubes, cleaning level 
has little to no effect on the derived velocity dispersion.  
When cleaned down to 1.5 times rms, non-residual scaled cubes give 
identical velocity dispersion values as residual-scaled cubes. 
Blanking seems to affect only the velocity dispersion of the Natural non-residual scaled cubes 
cleaned to 3.5 times rms. 
The effects of cleaning level, residual-scaling and blanking on the shapes of the super profiles 
can be seen in Figure~\ref{fig:clean_prof}.   
\section{Which type of data cube should be used for velocity dispersion analysis?}\label{sec:2}
In the following, we test which cubes (residual vs non-residual scaled) 
best recover an input H\,{\sc i} velocity dispersion using simulated data cubes. 
Using the AIPS task UVMOD, we insert a model in the $(u, v)$ data of NGC 3184. 
This model consists of a disk with a uniform, constant column density, a diameter of 150$\arcsec$ and a 
Gaussian velocity distribution. All profiles in the disk have the 
same mean velocity and velocity width. The model was inserted in an 
otherwise empty (i.e., containing no line-signal) part of the NGC 3184 $(u, v)$ 
data set. This data set was then Fourier-transformed and CLEANed following 
the THINGS data reduction procedure. As the model is embedded in the actual, 
observed $(u, v)$ data, the noise level in the resulting cube is identical to 
that in the original THINGS cube. For the NGC 3184 data used here the value is 
0.36 mJy beam$^{-1}$. We stop at the cleaning level adopted by THINGS 
\citep[2.5 times rms;][]{walter08} and we use the Natural weighting scheme. 
We create residual and non-residual scaled version of the simulated cubes. 
Note that the model spectra are already lined-up in velocity; therefore, in constructing 
the super profiles, no shifting of individual profiles was needed. We simply extracted 
the profiles at the location of the model disk and summed them. For the model 
residual-scaled cubes, no additional manual blanking of channel maps was needed. 
As the profiles are already lined up at known velocities,
we simply only consider the velocity range where the profiles are visible.

\subsection{Effects of residual scaling on second moment values and single Gaussian dispersion} 
In this section, we quantify the effects of low level uncleaned emission on 
the derivation of H\,{\sc i} velocity dispersion using second moment calculation 
and single Gaussian fits of H\,{\sc i} spectra. Here, the model cubes created using the above procedure 
all have Gaussian input spectra with a velocity dispersion 
of 12 $\rm{km~s^{-1}}$ but different peak flux values. 

\begin{figure*}
\centering
\includegraphics[scale=.23]{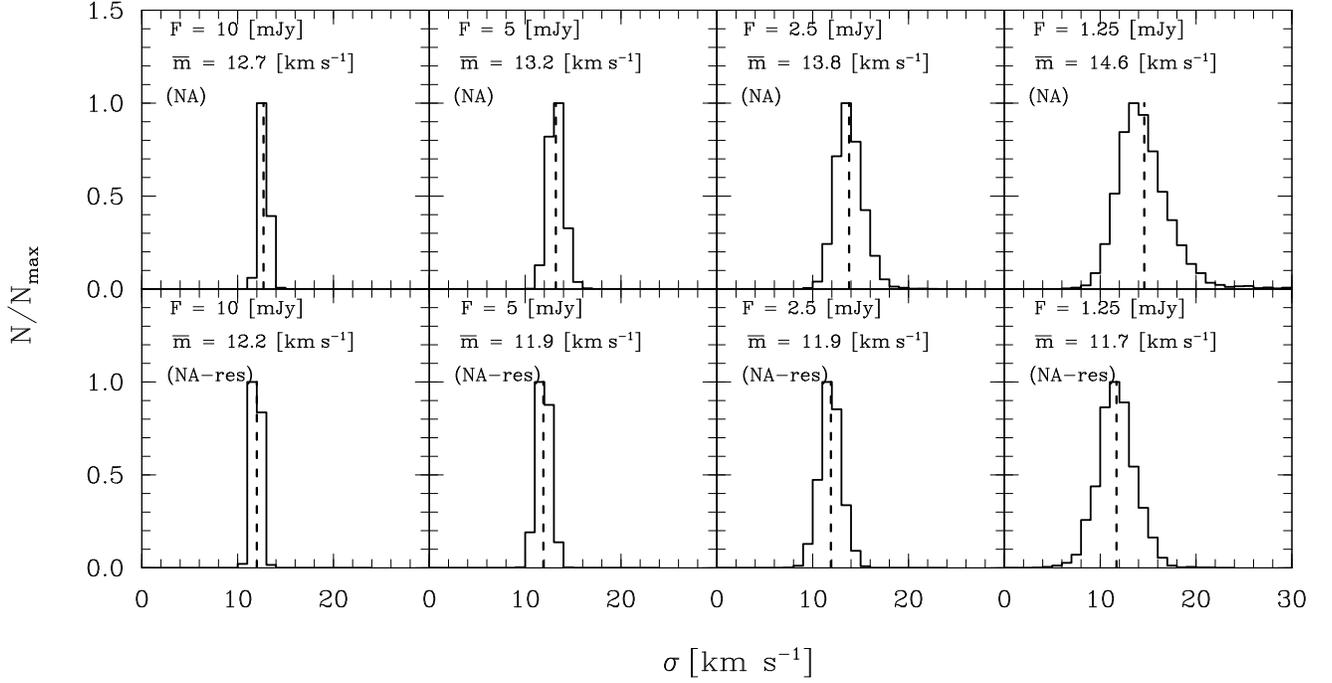}
\caption{Histograms of single Gaussian velocity dispersions from 
    natural non-residual (NA, first four panels) and natural residual-scaled (NA-res, last four panels) model data cubes. 
    $F$: input peak flux; $\bar{m}$: output mean velocity dispersion; \textit{Vertical dashed lines}: 
    input velocity dispersion (12 $\rm{km~s^{-1}}$) used to make the models.\\}
    \label{fig:hist_gau1}
\end{figure*}

\begin{figure*}
\centering
\includegraphics[scale=.23]{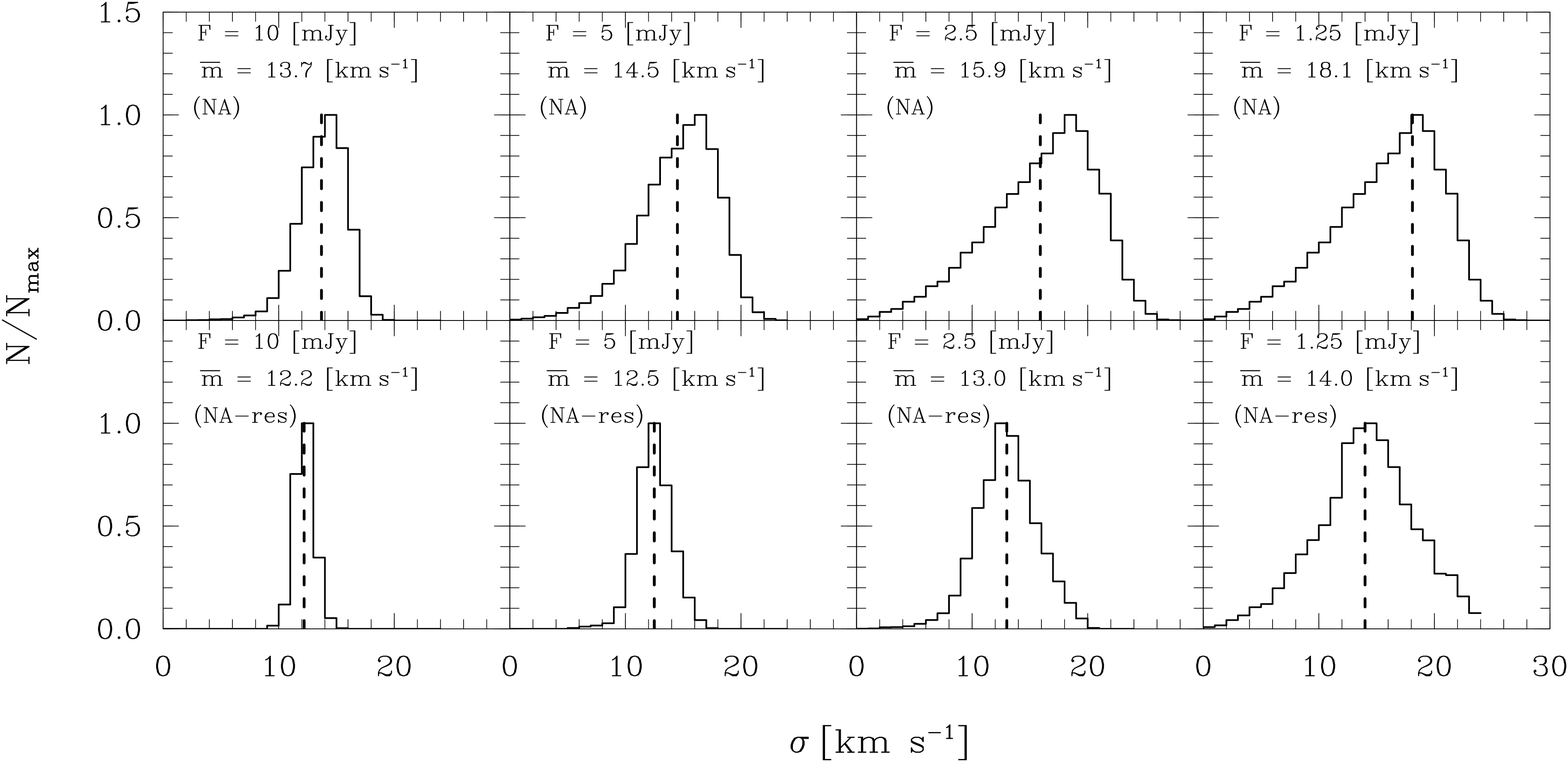}
\caption{Histograms of second moment values from 
    the natural non-residual scaled (NA, first four panels) and natural 
    residual-scaled (NA-res, last four panels) scaled model data cubes.  
    $F$: input peak flux; $\bar{m}$: output mean velocity dispersion; \textit{Vertical dashed lines}: 
    input velocity dispersion (12 $\rm{km~s^{-1}}$) used to make the models.}
   \label{fig:hist_mom2} 
\end{figure*}

Figure~\ref{fig:hist_gau1} and Figure~\ref{fig:hist_mom2} show histograms of the single Gaussian dispersion
and the mean second moment values derived from the model data cubes. We use the GIPSY task MOMENTS 
to calculate the second moment values. We do not apply any restrictions such as 
clipping of spectral channels to derive the second moment values. The model cubes do not 
contain any double profiles nor large noise spikes, so the values shown in Figure~\ref{fig:hist_gau1}
and Figure~\ref{fig:hist_mom2} are not affected by this. For the non-residual scaled cubes, 
the velocity dispersions from both the single Gaussian fits and the second moment maps depend on the input peak 
flux. The derived mean velocity dispersions tend to increase from higher to 
lower signal-to-noise (S/N). For the lowest input 
peak flux, the mean single Gaussian dispersion and second moment values differ from the input dispersion 
by $\sim$18\% and $\sim$34\%, respectively. For the residual-scaled model data cubes, 
the mean single Gaussian velocity dispersions are very similar to the input dispersion. We do, however, 
observe a steady increase in the mean second moment values as we decrease the 
input peak flux values. The derived value differs from the input dispersion by $\sim14$\% at the 
lowest input peak flux. The histograms of the second moment values have a broader distribution compared to those of the 
single Gaussian dispersions, especially for the non-residual scaled cubes at low S/N. 
Thus, second moment values are more sensitive to S/N than single Gaussian dispersions. 
The effects of S/N is more pronounced for the non-residual-scaled than 
for the residual-scaled cubes. The results from the models show that for the 
standard THINGS cubes (i.e., cleaned to 2.5 time the rms noise), 
residual-scaled cubes give more accurate velocity dispersion as opposed to 
non-residual scaled cubes. The latter tend to overestimate the velocity dispersions, especially in the low S/N regime. 

\subsection{Effects of residual scaling on velocity dispersions derived from a fitted two-Gaussian model}  
\begin{figure}[H]
    \centering
     \includegraphics[width=1.0\textwidth]{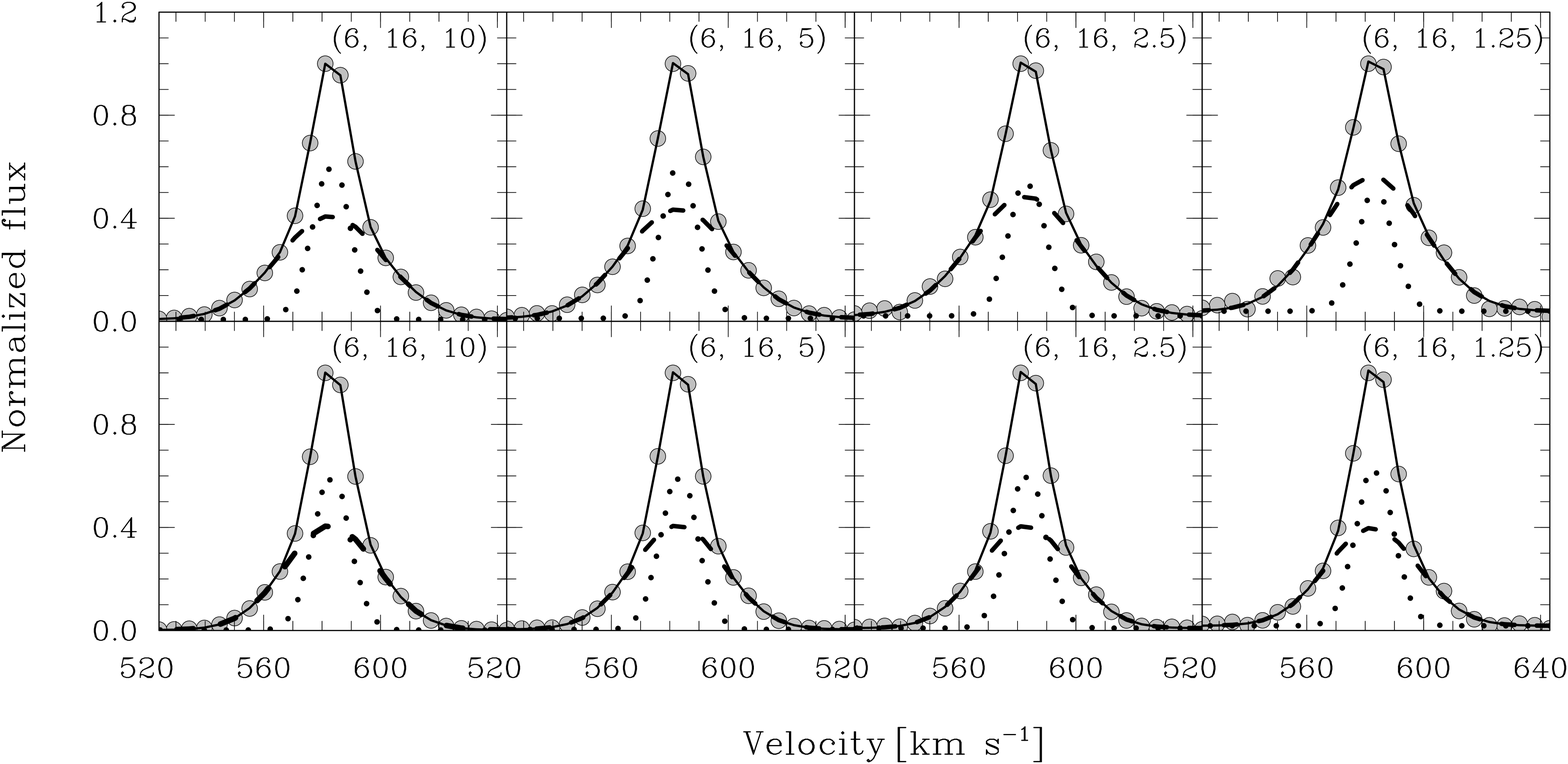} 
    \caption{\mbox{Synthetic super profiles from non-residual scaled cubes (top panels) and residual-scaled cubes 
    (bottom panels)}.\\ \mbox{The number in each plots are the input $\sigma_{n}$ ($\rm{km~s^{-1}}$), 
    $\sigma_{b}$ ($\rm{km~s^{-1}}$), and peak flux (i.e., $a_{n}$ + $a_{b}$ in mJy) values.}}
    \label{fig:models} 
\end{figure}

Here the model data cubes are constructed using narrow and 
broad Gaussian input spectra, each with different input profile 
parameter values ($\sigma_{n}$, $\sigma_{b}$, $A_{n}/A_{b}$, and the ratio of the 
narrow and broad component amplitudes, $a_{n}/a_{b}$). 
Tables \ref{tab:model_set1} illustrates the results from the various models where the input 
parameters are shown in bold. The conclusions can be summarized as follows. The residual and non-residual 
scaled model data cubes give similar $\sigma_{n}$ values, approaching the input values for a range of 
signal-to-noise (S/N). However, the non-residual scaled cubes 
give $\sigma_{b}$ values that are overestimated by up to $\sim$ 16\%, whereas the residual-scaled ones 
result in $\sigma_{b}$ values similar to the input values. In addition, the non-residual scaled cubes 
tend to underestimate the $A_{n}/A_{b}$ ratios, especially for lower input peak flux values (i.e., lower S/N). 
The derived $A_{n}/A_{b}$ values from the residual-scaled cubes, however, are similar to the input 
values and do not depend on the input peak flux. For the non-residual scaled cubes, 
when we go from higher to 
lower input peak flux, the $\sigma_{n}$ and $\sigma_{b}$ values 
stay basically the same, whereas the broad component amplitude becomes higher and therefore its area 
also gets larger. This results in a decrease in the $a_{n}/a_{b}$ and $A_{n}/A_{b}$ ratios. Thus in 
the low S/N regime, the $A_{n}/A_{b}$ ratio from non-residual scaled cubes can be underestimated. 
For example, in our model, where the S/N is $\sim$3.6, $A_{n}/A_{b}$ is underestimated by up to 
$\sim$48\%. For the residual-scaled cubes, the derived $A_{n}/A_{b}$ ratios and the input values agree 
within $\sim$18\% for the same S/N. To illustrate these, we show the shapes of the (overall) super profiles, 
the individual narrow and broad components in Figure~\ref{fig:models} for a range of S/N for both the residual and 
the non-residual scaled cubes. Figure~\ref{fig:models} clearly shows that 
the amplitudes of the fitted narrow and broad components 
from the non-residual scaled cubes are sensitive to the input S/N. However, the shapes of the fitted narrow/broad 
components from the residual-scaled cubes are not dependent on the input S/N.
\section{Discussion and Conclusion}\label{sec:discussion}
We have shown that, for observations where the shape of the dirty beam 
deviates significantly from that of a Gaussian, H\,{\sc i} velocity dispersion is sensitive to cleaning depth. 
If the dirty beam has prominent wings and if not very deeply cleaned, 
the cleaned map will consist of cleaned emission with a Gaussian beam sitting 
on a broad pedestal residual emission. Second moment values and broad component 
dispersions derived from such cubes will be overestimated as they are sensitive 
to profile wings. Blanking also affects the broad profile wings.
The narrow component flux 
and dispersion are, however, not affected by the dirty beam effects as they 
represent the core of the profile which is mostly composed of bright emission 
that is properly cleaned. The mismatch between the dirty beam and the fitted 
clean beam also leads to smaller $\sigma_{n}/\sigma_{b}$ and $A_{n}/A_{b}$ values. 
Analogous to the narrow component, the single Gaussian dispersions are less 
overestimated as the fitting routine tends to fit the core of the profiles.

\begin{figure*}
\centering
    \begin{tabular}{l}
        \includegraphics[scale=.2]{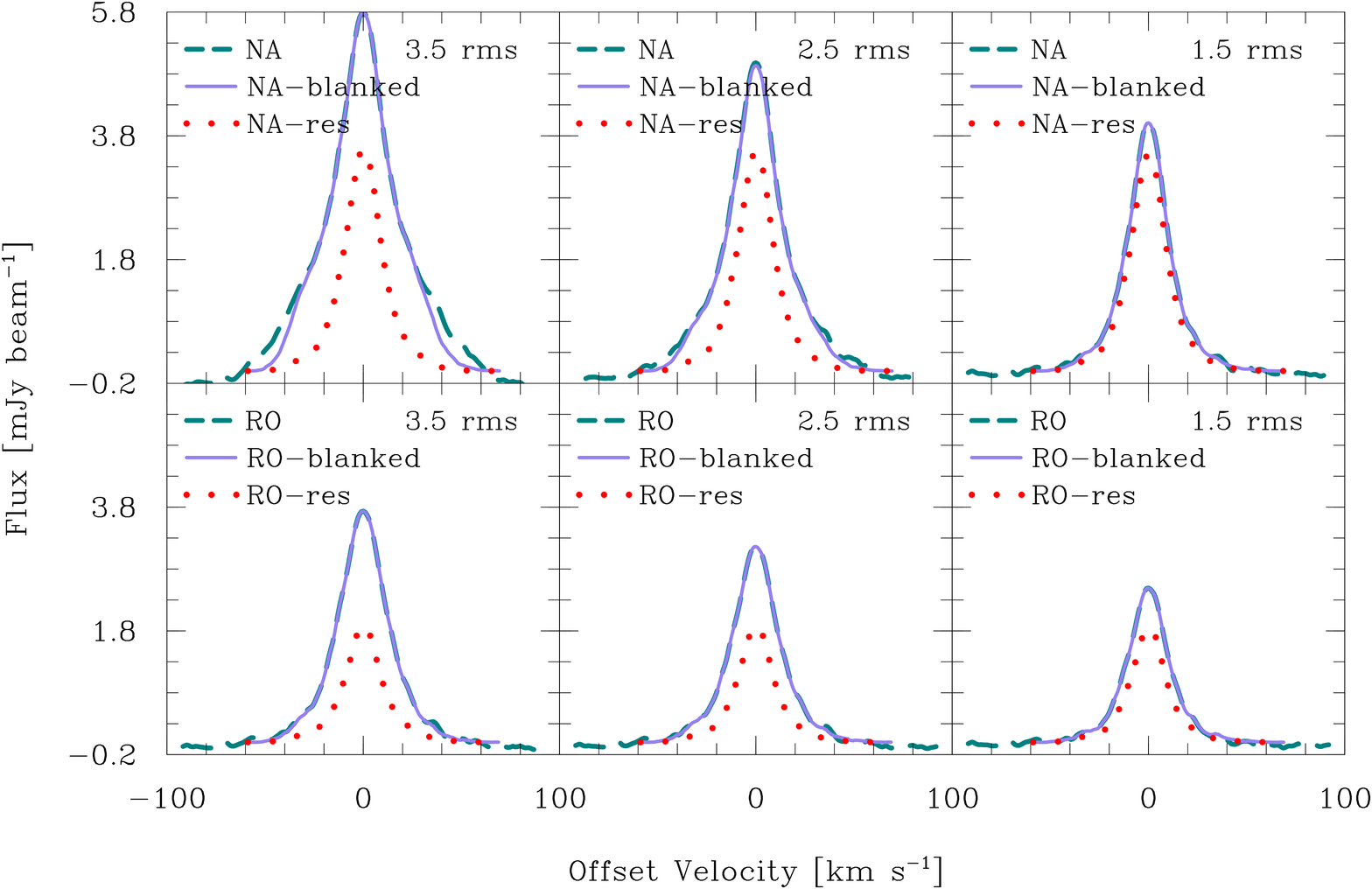}
    \end{tabular} 
    \caption{Super profile shapes from non-residual 
    scaled (\textit{green dashed lines}: not blanked; \textit{purple solid lines}: blanked) 
    and residual-scaled (red dotted lines) data cubes of NGC 3184 in annulus of 
    R $\sim$ 0.5 $\rm{r_{25}}$ and W $\sim$ 0.1 $\rm{r_{25}}$. 
    The top panels are profiles from 
    natural-weighted data cubes and the bottom panels are profiles from robust-weighted data cubes. 
    From left to right, the data cubes are cleaned down to a level of (3.5,  2.5, 1.5) times rms.}
    \label{fig:clean_prof}
 \end{figure*}

Robust weighting is less dependent on cleaning depth because the resulting 
dirty beam has less prominent wings. However, as seen in Section~\ref{sec:goal}, 
the Natural dirty beam can be substantially different from the fitted restoring beam. 
The effects of this mismatch on the estimate of H\,{\sc i} flux and velocity 
dispersion can be lowered by scaling the residuals by the ratio between the clean 
beam area and the dirty beam area \citep[see also][]{stilpetal13}. This is why the 
super profiles from residual-scaled cubes tend to have smaller wings than those 
from non-residual scaled cubes as seen in Figure~\ref{fig:clean_prof}. This implies 
that for arrays where the dirty beam is more Gaussian (e.g., the WSRT telescope) 
these effects are less prominent to absent. 
In conclusion, when dealing with observations where the dirty beam is 
significantly non-Gaussian (e.g., of multi-configuration VLA observations), 
we recommend the use of residual-scaled cubes or cubes that are cleaned 
close to the noise level (1.5 times the rms noise or deeper) for H\,{\sc i} velocity 
dispersion analysis. Cleaning to insufficient depth can lead to a significant 
misestimate of the velocity dispersion. While cleaning this deep is not always trivial 
for the classical CLEAN, it has been shown that 
the MSCLEAN algorithm does well in cleaning close to the noise 
level \citep{richetal08} without suffering from divergence. 
\floattable
{\renewcommand{\arraystretch}{1.1}
 \begin{deluxetable*}{c c c c c c c c c c c}
\tablecolumns{4}
\centering
\tablecaption{Parameters from different set of model data cubes I\label{tab:model_set1}}
\tablehead{
    \multicolumn{5}{c}{Non-residual scaled}&&\multicolumn{5}{c}{Residual-scaled}\\
    \cline{1-5}\cline{7-11}\\
    $\sigma_{n}$& $\sigma_{b}$&Peak flux&$a_{n}/a_{b}$& $A_{n}/A_{b}$ & & $\sigma_{n}$& $\sigma_{b}$&Peak 
  flux&$a_{n}/a_{b}$ & $A_{n}/A_{b}$\\
    $\rm{(km~s^{-1})}$& $\rm{(km~s^{-1})}$&(mJy)&&&& $\rm{(km~s^{-1})}$& $\rm{(km~s^{-1})}$&(mJy)& & \\
    1 & 2 & 3 & 4 & 5 & & 6 & 7 & 8 & 9 & 10
}
\startdata
\textbf{8.00} & \textbf{20.00} & \textbf{10.00} & \textbf{1.50} & \textbf{0.60} & & \textbf{8.00} & \textbf{20.00} & \textbf{10.00} & \textbf{1.50} & \textbf{0.60} \\ 
8.17 $\pm$ 0.06 & 22.19 $\pm$ 0.28 & 11.45 & 1.53 $\pm$ 0.03 & 0.56 $\pm$ 0.03 & & 7.98 $\pm$ 0.04 & 19.63 $\pm$ 0.16 & 11.12 & 1.51 $\pm$ 0.02 & 0.61 $\pm$ 0.03 \\ 
\textbf{8.00} & \textbf{20.00} & \textbf{5.00} & \textbf{1.50} & \textbf{0.60} & & \textbf{8.00} & \textbf{20.00} & \textbf{5.00} & \textbf{1.50} & \textbf{0.60} \\ 
8.32 $\pm$ 0.12 & 23.35 $\pm$ 0.57 & 5.99 & 1.44 $\pm$ 0.04 & 0.51 $\pm$ 0.05 & & 7.97 $\pm$ 0.09 & 19.33 $\pm$ 0.30 & 5.54 & 1.50 $\pm$ 0.05 & 0.62 $\pm$ 0.06 \\ 
\textbf{8.00} & \textbf{20.00} & \textbf{2.50} & \textbf{1.50} & \textbf{0.60} & & \textbf{8.00} & \textbf{20.00} & \textbf{2.50} & \textbf{1.50} & \textbf{0.60} \\ 
8.24 $\pm$ 0.25 & 23.48 $\pm$ 0.99 & 3.25 & 1.21 $\pm$ 0.07 & 0.42 $\pm$ 0.08 & & 7.94 $\pm$ 0.18 & 18.92 $\pm$ 0.60 & 2.75 & 1.48 $\pm$ 0.10 & 0.62 $\pm$ 0.12 \\ 
\textbf{8.00} & \textbf{20.00} & \textbf{10.00} & \textbf{2.33} & \textbf{0.93} & & \textbf{8.00} & \textbf{20.00} & \textbf{10.00} & \textbf{2.33} & \textbf{0.93} \\ 
8.15 $\pm$ 0.05 & 22.65 $\pm$ 0.35 & 11.45 & 2.29 $\pm$ 0.04 & 0.82 $\pm$ 0.05 & & 7.98 $\pm$ 0.04 & 19.50 $\pm$ 0.21 & 11.12 & 2.35 $\pm$ 0.04 & 0.96 $\pm$ 0.05 \\ 
\textbf{8.00} & \textbf{20.00} & \textbf{5.00} & \textbf{2.33} & \textbf{0.93} & & \textbf{8.00} & \textbf{20.00} & \textbf{5.00} & \textbf{2.33} & \textbf{0.93} \\ 
8.23 $\pm$ 0.10 & 23.77 $\pm$ 0.73 & 5.99 & 2.08 $\pm$ 0.06 & 0.72 $\pm$ 0.08 & & 7.98 $\pm$ 0.08 & 19.20 $\pm$ 0.41 & 5.54 & 2.35 $\pm$ 0.09 & 0.98 $\pm$ 0.11 \\ 
\textbf{8.00} & \textbf{20.00} & \textbf{2.50} & \textbf{2.33} & \textbf{0.93} & & \textbf{8.00} & \textbf{20.00} & \textbf{2.50} & \textbf{2.33} & \textbf{0.93} \\ 
8.19 $\pm$ 0.19 & 23.39 $\pm$ 1.10 & 3.25 & 1.69 $\pm$ 0.09 & 0.59 $\pm$ 0.11 & & 7.96 $\pm$ 0.15 & 18.80 $\pm$ 0.79 & 2.75 & 2.34 $\pm$ 0.18 & 0.99 $\pm$ 0.22 \\
&&&&&&&&&&\\
\hline\hline\\[-.25cm]
&\multicolumn{10}{c}{\small Parameters from different set of model data cubes II}\\
\cline{1-5}\cline{7-11}\\
\textbf{6.00} & \textbf{16.00} & \textbf{10.00} & \textbf{1.49} & \textbf{0.56} & & \textbf{6.00} & \textbf{16.00} & \textbf{10.00} & \textbf{1.49} & \textbf{0.56} \\ 
6.16 $\pm$ 0.04 & 17.89 $\pm$ 0.14 & 11.45 & 1.56 $\pm$ 0.02 & 0.54 $\pm$ 0.02 & & 6.00 $\pm$ 0.03 & 15.76 $\pm$ 0.10 & 11.06 & 1.56 $\pm$ 0.02 & 0.59 $\pm$ 0.02 \\ 
\textbf{6.00} & \textbf{16.00} & \textbf{5.00} & \textbf{1.49} & \textbf{0.56} & & \textbf{6.00} & \textbf{16.00} & \textbf{5.00} & \textbf{1.49} & \textbf{0.56} \\ 
6.20 $\pm$ 0.09 & 18.39 $\pm$ 0.33 & 5.99 & 1.42 $\pm$ 0.04 & 0.48 $\pm$ 0.04 & & 6.01 $\pm$ 0.06 & 15.61 $\pm$ 0.20 & 5.51 & 1.56 $\pm$ 0.04 & 0.60 $\pm$ 0.05 \\ 
\textbf{6.00} & \textbf{16.00} & \textbf{2.50} & \textbf{1.49} & \textbf{0.56} & & \textbf{6.00} & \textbf{16.00} & \textbf{2.50} & \textbf{1.49} & \textbf{0.56} \\ 
6.15 $\pm$ 0.22 & 18.45 $\pm$ 0.64 & 3.25 & 1.19 $\pm$ 0.07 & 0.40 $\pm$ 0.08 & & 6.04 $\pm$ 0.12 & 15.52 $\pm$ 0.38 & 2.73 & 1.60 $\pm$ 0.08 & 0.62 $\pm$ 0.09 \\ 
\textbf{6.00} & \textbf{16.00} & \textbf{1.25} & \textbf{1.49} & \textbf{0.56} & & \textbf{6.00} & \textbf{16.00} & \textbf{1.25} & \textbf{1.49} & \textbf{0.56} \\ 
5.97 $\pm$ 0.47 & 17.67 $\pm$ 1.01 & 1.84 & 0.92 $\pm$ 0.11 & 0.31 $\pm$ 0.12 & & 6.13 $\pm$ 0.23 & 15.66 $\pm$ 0.77 & 1.34 & 1.71 $\pm$ 0.17 & 0.67 $\pm$ 0.20 \\ 
\textbf{6.00} & \textbf{16.00} & \textbf{10.00} & \textbf{2.32} & \textbf{0.87} & & \textbf{6.00} & \textbf{16.00} & \textbf{10.00} & \textbf{2.32} & \textbf{0.87} \\ 
6.12 $\pm$ 0.04 & 18.15 $\pm$ 0.21 & 11.45 & 2.33 $\pm$ 0.04 & 0.78 $\pm$ 0.04 & & 6.00 $\pm$ 0.03 & 15.70 $\pm$ 0.13 & 11.06 & 2.43 $\pm$ 0.04 & 0.93 $\pm$ 0.04 \\ 
\textbf{6.00} & \textbf{16.00} & \textbf{5.00} & \textbf{2.32} & \textbf{0.87} & & \textbf{6.00} & \textbf{16.00} & \textbf{5.00} & \textbf{2.32} & \textbf{0.87} \\ 
6.16 $\pm$ 0.08 & 18.62 $\pm$ 0.42 & 5.99 & 2.06 $\pm$ 0.06 & 0.68 $\pm$ 0.07 & & 6.02 $\pm$ 0.05 & 15.59 $\pm$ 0.26 & 5.51 & 2.47 $\pm$ 0.08 & 0.95 $\pm$ 0.09 \\ 
\textbf{6.00} & \textbf{16.00} & \textbf{2.50} & \textbf{2.32} & \textbf{0.87} & & \textbf{6.00} & \textbf{16.00} & \textbf{2.50} & \textbf{2.32} & \textbf{0.87} \\ 
6.13 $\pm$ 0.19 & 18.29 $\pm$ 0.79 & 3.25 & 1.68 $\pm$ 0.11 & 0.56 $\pm$ 0.12 & & 6.06 $\pm$ 0.10 & 15.54 $\pm$ 0.50 & 2.73 & 2.56 $\pm$ 0.15 & 1.00 $\pm$ 0.17 \\ 
\textbf{6.00} & \textbf{16.00} & \textbf{1.25} & \textbf{2.32} & \textbf{0.87} & & \textbf{6.00} & \textbf{16.00} & \textbf{1.25} & \textbf{2.32} & \textbf{0.87} \\ 
6.01 $\pm$ 0.43 & 17.08 $\pm$ 1.23 & 1.84 & 1.28 $\pm$ 0.17 & 0.45 $\pm$ 0.19 & & 6.10 $\pm$ 0.19 & 15.67 $\pm$ 1.00 & 1.34 & 2.73 $\pm$ 0.32 & 1.06 $\pm$ 0.36 \\ 
\textbf{6.00} & \textbf{16.00} & \textbf{10.00} & \textbf{0.80} & \textbf{0.30} & & \textbf{6.00} & \textbf{16.00} & \textbf{10.00} & \textbf{0.80} & \textbf{0.30} \\ 
6.25 $\pm$ 0.07 & 17.59 $\pm$ 0.13 & 11.45 & 0.88 $\pm$ 0.01 & 0.31 $\pm$ 0.02 & & 5.99 $\pm$ 0.04 & 15.80 $\pm$ 0.08 & 11.06 & 0.82 $\pm$ 0.01 & 0.31 $\pm$ 0.01 \\ 
\textbf{6.00} & \textbf{16.00} & \textbf{5.00} & \textbf{0.80} & \textbf{0.30} & & \textbf{6.00} & \textbf{16.00} & \textbf{5.00} & \textbf{0.80} & \textbf{0.30} \\ 
6.33 $\pm$ 0.12 & 18.24 $\pm$ 0.24 & 5.99 & 0.83 $\pm$ 0.02 & 0.29 $\pm$ 0.03 & & 5.99 $\pm$ 0.09 & 15.67 $\pm$ 0.14 & 5.51 & 0.82 $\pm$ 0.02 & 0.31 $\pm$ 0.02 \\ 
\textbf{6.00} & \textbf{16.00} & \textbf{2.50} & \textbf{0.80} & \textbf{0.30} & & \textbf{6.00} & \textbf{16.00} & \textbf{2.50} & \textbf{0.80} & \textbf{0.30} \\ 
6.21 $\pm$ 0.26 & 18.55 $\pm$ 0.46 & 3.25 & 0.72 $\pm$ 0.04 & 0.24 $\pm$ 0.05 & & 6.02 $\pm$ 0.17 & 15.52 $\pm$ 0.27 & 2.73 & 0.82 $\pm$ 0.04 & 0.32 $\pm$ 0.05 \\ 
\textbf{6.00} & \textbf{16.00} & \textbf{1.25} & \textbf{0.80} & \textbf{0.30} & & \textbf{6.00} & \textbf{16.00} & \textbf{1.25} & \textbf{0.80} & \textbf{0.30} \\ 
5.91 $\pm$ 0.57 & 18.14 $\pm$ 0.76 & 1.84 & 0.55 $\pm$ 0.06 & 0.18 $\pm$ 0.07 & & 6.13 $\pm$ 0.32 & 15.52 $\pm$ 0.54 & 1.34 & 0.86 $\pm$ 0.08 & 0.34 $\pm$ 0.10 \\ 
\enddata
\tablecomments{The input parameters for each models are represented in bold, whereas the derived 
  parameters from the resulting cubes are shown in regular fonts. 
  Column 1 \& 6: velocity dispersion of the narrow component; Column 2 \& 7: velocity dispersion of the broad component; 
  Column 3 \& 8: input peak flux values; 
  Column 4 \& 9: ratio between the input amplitude of the narrow and the broad components.  
  Column 5 \& 10: flux ratio of the narrow component and the broad component. }
\end{deluxetable*}
}   


\clearpage
\onecolumngrid
\twocolumngrid
\acknowledgments
We thank the anonymous referee for useful and constructive comments. 
We are also grateful to Dr. Fabian Walter for useful discussions during the preparation of this manuscript. 

R.I. acknowledges funding from the Alexander von Humboldt foundation through the Georg Forster Research Fellowship 
programme. 

R.I. working visit to the Netherlands was supported by the South 
African National Research Foundation-Netherlands Organisation for Scientific Research (NRF-NWO) 
exchange programme in ``Astronomy, and Enabling Technologies for Astronomy''. 

The work of W.J.G.d.B. was supported by the European Commission (grant FP7-PEOPLE-2012-CIG \#333939).

\end{document}